\documentclass[10pt,journal,compsoc]{IEEEtran}
\ifCLASSOPTIONcompsoc
  \usepackage[noadjust]{cite}
  
\else
  \usepackage{cite}
\fi
\ifCLASSINFOpdf
   \usepackage[pdftex]{graphicx}
   \graphicspath{{figures/}{pictures/}{images/}{./}}
   \DeclareGraphicsExtensions{.pdf,.jpeg,.png}
\else
\fi
\usepackage{dblfloatfix} 

\usepackage{tabu}                      
\usepackage{booktabs}                  

\usepackage{amssymb}                   
\usepackage{amsfonts}                  
\usepackage{amsmath}

\usepackage{caption}
\usepackage{subcaption}

\usepackage{listings}               

\usepackage{breqn}                    
\usepackage[ruled,vlined, linesnumbered]{algorithm2e}  

\usepackage{bm}
\newcommand{\vect}[1]{\bm{#1}} 
\usepackage{tikz} 
\newcommand{\mycbox}[1]{\tikz{\path[draw=black,fill={#1}] (0.01,0.01) rectangle (0.17cm,0.17cm);}}

\definecolor{color1}{HTML}{FFD966}
\definecolor{color2}{HTML}{D9D9D9}

\usepackage{enumitem} 

\usepackage{xcolor}

\newcommand{\removed}[1]{\textcolor{red}{\sout{#1}}}

\def \cleanversion{} 
\ifx\cleanversion\undefined
\else

  \renewcommand{\removed}[1]{\iffalse#1\fi}
\fi

\begin{document}
%
\title{Reinforcement Learning for \\ Load-balanced Parallel Particle Tracing}
%
%
%
%

\author{Jiayi~Xu, 
        Hanqi~Guo,~\IEEEmembership{Member,~IEEE,} 
        Han-Wei~Shen,~\IEEEmembership{Member,~IEEE,} 
        Mukund~Raj, 
        Skylar~W.~Wurster, 
        and~Tom~Peterka,~\IEEEmembership{Member,~IEEE}
\IEEEcompsocitemizethanks{\IEEEcompsocthanksitem Jiayi~Xu, Han-Wei~Shen, and Skylar~W.~Wurster are with the Department of Computer Science and Engineering, The Ohio State University, Columbus, OH, 43210, USA.\protect\\
E-mail: \{xu.2205, shen.94, wurster.18\}@osu.edu
\IEEEcompsocthanksitem Hanqi~Guo and Tom~Peterka are with the Mathematics and Computer Science Division, Argonne National Laboratory, Lemont, IL 60439, USA.\protect\\
E-mail: \{hguo, tpeterka\}@anl.gov
\IEEEcompsocthanksitem Mukund~Raj is with the Stanley Center for Psychiatric Research, Broad Institute of MIT and Harvard, Cambridge, MA 02142, USA.\protect\\
E-mail: mraj@broadinstitute.org}
}
\markboth{IEEE Transactions of Visualization and Computer Graphics,~Vol.~X, No.~X, September~2021}%
{Xu \MakeLowercase{\textit{et al.}}: Reinforcement Learning for Load-balanced Parallel Particle Tracing}
\IEEEtitleabstractindextext{%
\begin{abstract}
We explore an online reinforcement learning (RL) paradigm to dynamically optimize parallel particle tracing performance in distributed-memory systems. Our method combines three novel components: (1) a work donation algorithm, (2) a high-order workload estimation model, and (3) a communication cost model. First, we design an RL-based work donation algorithm. Our algorithm monitors workloads of processes and creates RL agents to donate data blocks and particles from high-workload processes to low-workload processes to minimize program execution time. The agents learn the donation strategy on the fly based on reward and cost functions designed to consider processes' workload changes and data transfer costs of donation actions. Second, we propose a workload estimation model, helping RL agents estimate the workload distribution of processes in future computations. Third, we design a communication cost model that considers both block and particle data exchange costs, helping RL agents make effective decisions with minimized communication costs. We demonstrate that our algorithm adapts to different flow behaviors in large-scale fluid dynamics, ocean, and weather simulation data. Our algorithm improves parallel particle tracing performance in terms of parallel efficiency, load balance, and costs of I/O and communication for evaluations with up to 16,384 processors. 
\end{abstract}

\begin{IEEEkeywords}
Distributed and parallel particle tracing, dynamic load balancing, reinforcement learning. 
\end{IEEEkeywords}}

\maketitle

\IEEEdisplaynontitleabstractindextext

%
\IEEEpeerreviewmaketitle

\IEEEraisesectionheading{\section{Introduction}\label{sec:introduction}}

%
%
%
%

\IEEEPARstart{A}{s} 
the size and complexity of vector-field data increase, distributed and parallel particle tracing becomes essential for visualizing and analyzing large-scale data from scientific simulations. 
For example, distributed texture-based flow visualizations, such as line integral convolution (LIC)~\cite{cabral1993imaging, muraki2003pc} and finite-time Lyapunov exponents (FTLEs)~\cite{haller2001distinguished, nouanesengsy2012parallel, zhang2017dynamic}, benefit from distributed particle advection for efficient computation of streamlines and pathlines. Other applications of distributed and parallel particle tracing also play essential roles in scientific data analysis, such as distributed streamsurface computation~\cite{lu2014scalable} and source-destination queries~\cite{kendall2011simplified, guo2014advection, zhang2017dynamic}, just to name a few.

According to existing studies~\cite{pugmire2009scalable, peterka2011study}, the scalability and performance of distributed particle tracing are highly dependent on two aspects: (1) the workload balance of parallel processes and (2) the cost of communications. 
First, the workload of processes in distributed and parallel particle tracing can be imbalanced. 
Uneven distributions of complex flow features (e.g., critical points and vortices) in space usually lead to uneven distributions of particle positions during the tracing. 
Second, the cost of interprocess communications can be high due to the exchange of particles or data blocks. For example, the circular flow patterns in input vector field data usually lead to frequent particle transfer among data blocks, causing a high communication cost. 

This work is motivated by the need to simultaneously maximize load balance and minimize communication costs to reduce execution time for parallel particle tracing. Two challenges exist to build such an optimization algorithm to optimize online performance with minimal overhead. First, such an algorithm must be able to balance workloads as much as possible while avoiding frequent and unnecessary data movements. Second, the optimization must be achieved in real-time so as not to slow down parallel particle tracing. To the best of our knowledge, solving these two challenges is still an open problem. 

We perform simultaneous optimization of workload balance and communication efficiency based on reinforcement learning (RL) studies. 
The optimization requires distributing data blocks and particles among processes to maximize the workload balance and minimize the communication overhead, which can be categorized into the integer programming problem and is NP-complete \cite{garey1974some, gary1979computers}. Additionally, the particle positions continue to change during the parallel execution, leading to volatile information for optimization decisions; hence, classic methods such as dynamic programming are ineffective. 
RL approaches are developed to create agents to adapt decision-making with learning from the dynamic environments and maximizing reward functions~\cite{sutton2018reinforcement}. Reward and cost functions are designed in this paper to incorporate both workload balance and communication costs. 

To address the aforementioned challenges, we introduce three components that work hand in hand to enable online performance optimization for parallel particle tracing in distributed-memory systems: (1) work donation algorithm, (2) workload estimation model, and (3) communication cost model. 
First, we propose an RL-based work donation algorithm to allow processes to balance their workloads periodically. 
We associate an RL agent with each process. An agent is trained and used to move works from processes with more workload to those with less workload. 
Rewards guide the agents' behaviors and are designed according to the distributions of workloads and costs of data transfer in order to create balanced workloads among processes with minimized costs. 
Second, we design an online and high-order workload estimation model, which can estimate blockwise particle advection integration steps and time based on historical data that is recorded during the run time. The model learns the historical data from particles' high-order data access patterns, and dynamically adapts to different flow behaviors. 
Third, we construct a communication cost model to estimate the data transfer time of both blocks and particles based on the historical data transfer since the beginning of the run, allowing the model to adapt to the available network bandwidth. The communication cost model is constructed based on a linear transmission model~\cite{kielmann2001network, chan2007collective, traff2019optimal} that models the cost to be a constant latency plus time proportional to the data size.


To manage and optimize the activities across processes, we orchestrate the pipeline of the workload-balanced parallel particle tracing by taking the communication costs into account. 
First, we decompose the entire input vector field domain into data blocks, which are then assigned to participating parallel processes. We distribute particle seeds and trace the particles within the data blocks. 
Second, when particle tracing runs in parallel, processes collect statistical information of particle advection and data transfer to establish the workload estimation model and communication cost model. The estimated costs of future computations and communications are passed to processes' RL agents for agents to make accurate decisions. 
Third, agents of processes transfer data blocks and particles among processes to balance workloads with minimized data transfer costs. 
After agents take actions, feedback rewards considering workload distributions and data transfer costs are computed to improve agents' decision-making ability for future decisions. 
We tailor a policy-gradient-based reinforcement learning algorithm to train the agents and efficiently make load balancing decisions. 

We evaluate our technique with applications from fluid dynamics, ocean, and weather. We run our prototype implementation on a supercomputer with up to $16,384$ processors. Our method outperforms the state-of-the-art work stealing/requesting in terms of parallel efficiency, load balance, and the cost of I/O and communication. The contributions of this paper are threefold: 
\begin{enumerate} 
\item We propose a reinforcement learning based work donation algorithm for distributed-memory systems to optimize load balance with minimized communication costs and dynamically adapt to flow behaviors and available network bandwidth. 
\item We design a high-order workload estimation model to predict blockwise particle advection workloads. 
\item We introduce the use of a linear transmission model to estimate interprocess communications' costs. 
\end{enumerate}

\section{Related Works}
\label{sect:rl_related_work}

We summarize related works on parallel particle tracing in distributed-memory systems. In general, parallel particle tracing can be categorized into distributed-memory~\cite{sujudi1996integration, yu2007parallel, chen2008optimizing, pugmire2009scalable, peterka2011study, nouanesengsy2011load, kendall2011simplified, muller2013distributed, guo2013coupled, lu2014scalable, guo2014scalable, zhang2017dynamic, zhang2018dynamic, binyahib2019lifeline, camp2013gpu, childs2014particle} and shared-memory~\cite{lane1994ufat, lane1995parallelizing, cabral1995highly, camp2010streamline, pugmire2018performance, schwartz2021machine} settings, where our paper is focused on the distributed particle tracing; the former focuses on computations in independent processes with separate memory spaces, and the latter is done in computing environments that share the same memory space, including many- and multi-core processors. 
In distributed-memory settings, two strategies exist, including data-parallel
and task-parallel. We refer to literature~\cite{pugmire2012parallel, zhang2018survey} for a comprehensive review of parallel particle tracing.

\subsection{Load Balancing}




There are two basic strategies: (1) static and (2) dynamic load balancing, where this paper studies the dynamic one.

\textbf{Static load balancing:}
The data partition is predetermined and optimized before parallel particle tracing is performed. 
Peterka et al.~\cite{peterka2011study} used a static round-robin 
strategy to assign data blocks to parallel processes to balance workload. 
Alternatively, Nouanesengsy et al. \cite{nouanesengsy2011load} proposed a matrix-based optimization algorithm to assign blocks to processes for workload balancing. 

\textbf{Dynamic load balancing:}
Workloads of processes are periodically optimized. Existing dynamic load balancing algorithms have three categories: (1) domain (re-)partitioning, (2) master/slave, and (3) work stealing/requesting. 
First, Peterka et al. \cite{peterka2011study} applied the recursive coordinate bisection (RCB)~\cite{berger1987partitioning} to partition data dynamically to make each process have a similar estimated workload. 
Zhang et al.~\cite{zhang2017dynamic, zhang2018dynamic} improved the RCB based method using a constrained k-d tree~\cite{zhang2017dynamic} and a workload prediction model~\cite{zhang2018dynamic}. 
Second, Pugmire et al.~\cite{pugmire2009scalable} proposed a master/slave based algorithm, where master processes dynamically move particles to idle slaves when slave processes have no work to do. 
Third, M{\"u}ller et al.~\cite{muller2013distributed} and Lu et al.~\cite{lu2014scalable} applied work stealing/requesting scheme~\cite{blumofe1999scheduling, dinan2009scalable} for distributed and parallel particle tracing, where idle processes repeatedly steal particles from random processes to reduce the total idle time. 
Binyahib et al.~\cite{binyahib2019lifeline} applied the lifeline technique~\cite{saraswat2011lifeline} to connect processes with a one-bit difference in process ranks so that particles can be redistributed among connected processes after random stealing fails.

\subsection{Workload Estimation}

Workload estimation is usually used to assist in balancing workloads among processes. Existing estimation strategies have two categories: (1) blockwise workload estimation and (2) particle-wise workload estimation, where this paper studies the blockwise workload estimation in~Section~\ref{sect:workload_estimation}.

\textbf{Blockwise workload estimation:}
One may estimate each data block's workload and assign data blocks to processes to have a balanced estimated workload. Nouanesengsy et al.~\cite{nouanesengsy2011load} advected a set of uniformly seeded particles in a preprocessing stage to determine the flow characteristics of each vector-field data block. 
Peterka et al.~\cite{peterka2011study} used the historical number of advection steps per particle within a data block to estimate the workload of incoming particles of that block as the future workload of that block.  


\textbf{Particle-wise workload estimation:}
One may estimate each particle's workload from the current time to the tracing termination and assign particles with a similar estimated workload to each process. 
One may assume each particle requires similar computation time and balance workloads of processes by balancing particle counts~\cite{pugmire2009scalable, muller2013distributed, lu2014scalable}. 
Zhang et al.~\cite{zhang2018dynamic} estimated the workload of each particle from the particle exits the current accessed data block to advection completion, with the construction of access dependency graphs (or, flow graphs)~\cite{xu2010flow, chen2011flow, chen2012flow}.

\subsection{Input/Output}
I/O for data block loading is a common bottleneck for the scalability and performance of distributed and parallel particle tracing methods. Two types of techniques are proposed to reduce the I/O cost for loading and fetching data blocks in parallel particle tracing. 

\textbf{On-demand data loading:}
Data blocks are loaded from disks as long as no more work can be completed based on the blocks in memory. 
Pugmire et al.~\cite{pugmire2009scalable} proposed the on-demand data loading strategy, which is followed by various parallel particle tracing studies~\cite{camp2010streamline, lu2014scalable}. 


\textbf{Prefetching:}
Prefetching data blocks at each disk access with knowledge of data access dependencies can reduce I/O costs. 
Chen et al.~\cite{chen2012flow} constructed an access dependency graph~\cite{chen2011flow} to organize data blocks with strong access dependencies closely in disk file layout to reduce disk seek time, and proposed an out-of-core method to prefetch consecutive data blocks with access dependencies for efficient pathline computation. 
The work of Guo et al.~\cite{guo2014advection} builds hint graphs to record the data access dependencies between fine-grained blocks and prefetches data blocks with dependant access patterns into a parallel key-value store to reduce the latency of data access for unsteady flow visualizations. 
Zhang et al.~\cite{zhang2016efficient} studied Markov-chain based high-order access dependencies in unsteady flow fields, enabling high-order data prefetching for the computation of pathlines and acquiring improved prefetching accuracy. 
Hong et al.~\cite{hong2018access} studied predicting data accesses using Long Short-Term Memory (LSTM) models for data block prefetching.

\begin{figure*}[tb]
  \centering 

  \includegraphics[width=\linewidth]{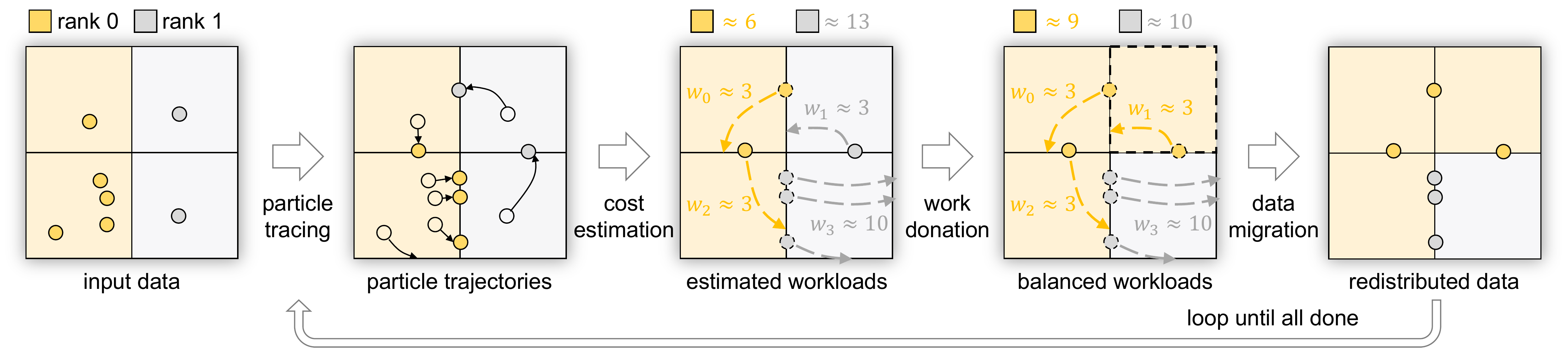}
  \caption{
  A schematic diagram of our RL based load-balanced parallel particle tracing. Two processes with rank 0 and 1 are colored in orange \protect\mycbox{color1} and gray \protect\mycbox{color2}, respectively. We use $w_i$ to indicate block $i$'s estimated advection workloads in seconds. The two processes' estimated advection workloads are labeled at the top, which are the sums of the owned blocks' workloads. 
  }
  \label{fig:schematic_diagram}
\end{figure*}




\section{Algorithm Overview}
\label{sect:rl_alg_overview}



Our algorithm aims to adjust the assignment of data blocks among processes to minimize the total execution time of parallel particle tracing.

\subsection{Optimization Problem Statement}


We seek to obtain a data block assignment $\mathcal{B}_t$ dependent on execution time $t$, such that to minimize the future execution time after time $t$, where $\mathcal{B}_t=(B_{t,0}, B_{t,1}, ... , B_{t, n_p-1})$. $B_{t, l}$ denotes the block set assigned to process $l$ (i.e., the process with rank $l$) at time $t$, and $n_p$ is number of total participating processes. The block sets satisfy that $B_{t,0} \cup B_{t,1} \cup ... \cup B_{t, n_p-1}=B$, where $B$ being the entire set of data blocks, and $B_{t, l} \cap B_{t, l^\prime} = \emptyset$ for $l\neq l^\prime$. We require that different processes do not share the same data block for four reasons. First of all, this helps constrain the solution space and make the optimization problem tractable. 
Second, we can efficiently maintain a distributed block-to-process mapping as a vector representation using \texttt{MPI} one-sided communications with low overhead. 
Third, this can reduce the total memory requirement. 
Fourth, this is a typical practice, as shown in several existing papers~\cite{peterka2011study, lu2014scalable}. 

Given the execution time of parallel particle tracing can be abstracted into two components: computation time for particle advection and communication time for data transfer, we consider decomposing the execution time minimization objective into two requirements: 
\begin{description}[font={\normalfont\itshape}]
  \item[R1 Minimizing advection computation time:]
  Advection time is bounded by the process with the highest computation time. When other processes complete their work, they become idle and wait for the process with the most computation to complete. 
  We can balance workloads of processes and minimize the workload of the process with the maximal computation time to speed up the execution, where the \textit{advection workload of a process} is defined to be the time of particle advection in its blocks. 

  \item[R2 Minimizing communication cost:]
        Data entities, including blocks and particles, are exchanged between processes through communications. Hence, R2 can be decomposed into two sub-requirements. 
  \begin{description}[font={\normalfont\itshape}]
    \item[R2.1 Minimizing block transfer cost:]
        Balancing processes' workloads via exchanging data blocks incurs additional overhead, which we seek to minimize.
    \item[R2.2 Minimizing particle transfer cost:]
        The particles, that are advected out of their current blocks, have to be transferred to the processes own their next blocks, which incurs this cost we seek to minimize. 
  \end{description}
\end{description}
This optimization problem can be categorized as integer programming, which is NP-complete~\cite{garey1974some, gary1979computers}. The solution space is too huge to be solved by a polynomial-time algorithm. Hence, we use an RL-based algorithm to approximate the optimal solution of the dynamic load balancing problem.

  





\subsection{Algorithm Pipeline}


Our algorithm illustrated in Fig.~\ref{fig:schematic_diagram} builds on top of a data-parallel particle tracing pipeline. 

\textbf{Traditional data-parallel particle tracing pipeline: }
The input includes vector field data and a set of particle seeds. The output consists of particles' trajectories. The pipeline has three stages: (1) initialization, (2) blockwise particle tracing computation, and (3) termination detection. 

In the first initialization stage, we distribute the input vector field data and seeds among processes. We decompose the vector field domain into axis-aligned data blocks with similar sizes where the number of the blocks is higher than the number of processes; 
specifically for unsteady flow, a time-dependent vector field is decomposed into spacetime blocks following previous works~\cite{chen2012flow, zhang2016efficient, zhang2017dynamic, zhang2018dynamic, hong2018access}, with time being considered as an additional dimension to space following~\cite{tricoche2002topology, garth2004tracking}. 
We assign the blocks to the parallel processes using the static round-robin assignment~\cite{peterka2011study}. Each process is assigned with the same number of blocks. We then distribute the input particle seeds among processes owning corresponding blocks.


In the second stage, we perform blockwise particle tracing. For every \textit{round} of particle tracing, we advect particles within the data blocks until all particles either stop prematurely due to hitting critical points in steady flows or go out of current block boundaries. At this point, blocks are exchanged if necessary. Then the particles are sent to the processes owning the next blocks that they are entering before a new round of particle tracing is performed.  


Third, the particle tracing is repeated and terminates until all the particles go out of the global domain boundary or exceed the maximum advection steps. 

In the following, we introduce our dynamic load balancing scheme into the above pipeline that runs after each round of blockwise particle tracing. 

\textbf{RL-based dynamic load balancing pipeline:}
The input consists of the distribution of blocks and particles among processes. The output is the redistribution of blocks and particles among processes to have more balanced workloads and a lower execution time. The pipeline has three stages: (1) cost estimation, (2) work donation, and (3) data migration. 

First, we estimate processes' workloads (Section~\ref{sect:workload_estimation}) and communication costs (Section~\ref{sect:communication_cost}). 
A workload estimation model is constructed to estimate blockwise workloads for the next round of computation. When particles go out of their current blocks' boundaries, processes first exchange the counts of particles that will be transferred to the other processes with adjacent blocks. Then, we use the counts of incoming new particles to each block to estimate the workloads at the next round. 
A communication cost model is constructed to estimate the costs of transferring data blocks and transferring particles. 


Second, an RL-based work donation optimizer (Section~\ref{sect:reinforcement_learning}) is built to learn on the fly to optimize the workload balance of processes for the subsequent round of computation with minimized communication costs. Donors (processes with high estimated workloads) make donation requests for offloading works to receivers (processes with low workloads). 

Third, donors redistribute data blocks if receivers accept the donation requests. After receivers receive new data blocks, particles are transferred to their next corresponding processes and start another round of advection. 


\textbf{Solution for cold start problem:}
To prevent the cold start of the estimation models, we release seeds in multiple rounds as follows. 
Each process splits the initially-assigned seeds into ten batches uniformly, and releases one batch in the first round of the parallel tracing. One of the remaining batches will be released in the next round after the particles seeded in the previous round are advected out of their seeding blocks. This seeding process continues until all the ten batches of seeds are released. Hence, the computations for the particles seeded after the first round can benefit from the historical records saved from the previous rounds.

\textbf{Algorithm design considerations:} 
We design a \textit{model-based} RL algorithm; that is, we explicitly model distributed-computing environments and establish estimation models to predict costs of future computation and communication events in the environments. 
The typical pipeline of the model-based RL algorithms, as shown in~\cite{atkeson1997comparison, polydoros2017survey, kaiser2019model, pal2020brief}, is to (1) establish estimation models of environments based on real observations and (2) plan decision-making policies depending on the established models without the need for interacting with the actual environments, where the policies are improved to maximize rewards simulated from the established models of the environments. Compared with \textit{model-free} RL algorithms that do not model the environments and optimize policies directly using actual measurements, model-based RL algorithms usually have the following benefits: (1) higher sample efficiency (i.e., requiring fewer data samples for training), (2) better generality and transferability, and (3) better explainability due to explicit modeling of environments, as demonstrated in existing surveys~\cite{polydoros2017survey, moerland2020model, pal2020brief}. 
Unique in parallel particle tracing, there are two additional benefits for the optimizations based on the established models of distributed computing environments. 
First, we can make predictions and optimizations on situations that do not occur, such as the reward computation of rejected donations. 
Second, the training of policy functions is accelerated. The update of policies is decoupled from actual measurements and only depends on the established models of the environments; hence it does not need to wait for the related processes to finish all computation and communication tasks in the next round to provide the training data. This accelerates the policy function training because some particles' advection may take quite a long time in a single round.

\section{Work Donation Algorithm} \label{sect:reinforcement_learning}


We propose a reinforcement learning based work donation algorithm to minimize the total execution time of parallel particle tracing in distributed-memory systems. Our algorithm features (1) balancing processes' workloads with minimized communication costs and (2) proactively instructing overloaded processes to donate workloads to underloaded processes to minimize processes' idle time. 
Because processes' actual workloads are unknown before particles are advected, workload estimates are computed and then used as the input to our algorithm before particle advection takes place. 
Agents monitor workload distributions among processes, and are trained by taking rewards describing load imbalance improvement and execution cost reduction after making donations. 
From the training, agents learn policy strategies to maximize gained rewards to improve load imbalance among processes and minimize execution time. 

We explain why agents need to learn strategies for work donations. The straightforward strategy, high-workload processes always donating work to processes with the lowest workload, is highly possible to make donation receivers become overburdened after the donations are made. Also, it is challenging to consider different communication costs and coordinate processes' actions when donations are planned. 
To pursue a proper strategy for redistributing processes' works, an agent is created on each process to learn and adapt its action according to rewards obtained from historical donations. 

\begin{figure}[htb]
\centering

  \includegraphics[width=\columnwidth]{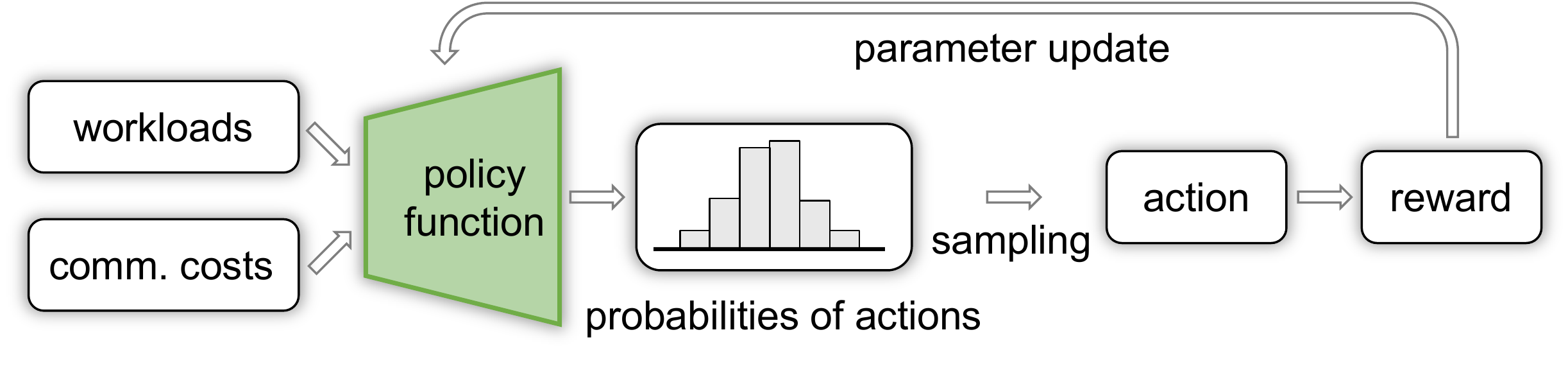}

  \caption{Every agent's decision-making and training pipeline. 
  }

  \label{fig:decision_making_diagram}
\end{figure}

We illustrate each agent's decision-making and training pipeline in~Fig.~\ref{fig:decision_making_diagram}. The agent of each donor uniformly samples a block from local blocks with non-zero workloads. By considering friend processes' workloads and different actions' communication costs, the agent computes probabilities of actions from the current policy function and then samples an action to determine whether the block should be donated to an underloaded process or not. Donors send donation requests to donation receivers. Receivers first assume all donations will be accepted, and send feedback with their current workloads plus all requested donations for donors to compute rewards and train their agents' policies. Note that if donors' donations would make receivers overloaded and cause more severe imbalanced workloads, the rewards will be negative, and the corresponding actions will be penalized in agents' training. If too many donation requests come simultaneously, the donation receivers can choose to reject certain donation requests, ensuring (1) the receivers' workloads would not be higher than the donors' after taking donations and (2) the receivers' local memory capacity is sufficient to hold the donations; the donation rejection mechanism also helps prevent obviously harmful donations when agents are not well-trained in early rounds. 

\subsection{Policy Gradient based Reinforcement Learning}
\label{sect:rl_model}

We create a multi-agent reinforcement learning based optimizer to maximize reward functions. 

\textbf{Preliminary: }
We first give a background of policy gradient based RL, which usually converges fast in applications~\cite{sutton2018reinforcement}. The agents' policy (i.e., probabilities of actions) are updated iteratively guided by reward and cost functions to find an optimal decision-making strategy. 

The policy-gradient-based methods use a Markov Decision Process (MDP) framework, which can be formulated by a tuple $(S, A, \pi_{\vect{\theta}}, R)$, where $S$ is a finite set of states, $A$ is a finite set of actions, $\pi_{\vect{\theta}}(a|s)$ is the policy function and denotes the probability taking an action $a \in A$ based on the current state $s \in S$, and $R(s, a, s^\prime)$ is the reward function based on the state $s$, the action $a$, and the new state $s^\prime$ after the action is taken. 

The policy gradient based methods parameterize $\pi_{\vect{\theta}}$, the policy of agents, using a parameter $\vect{\theta}$. The policy is optimized by updating $\vect{\theta}$ iteratively using gradient ascent to maximize rewards:  
\begin{equation} 
\vect{\theta} = \vect{\theta} + \alpha \cdot \nabla_{\vect{\theta}}E_{\vect{\theta}}[R(s, a, s^\prime)], 
\end{equation}
where $\alpha$ is the learning rate controlling how quickly the policy parameter is adapted to the rewards. 
$E_{\vect{\theta}}[R(s, a, s^\prime)]$ is the expectation of rewards, and $\nabla_{\vect{\theta}}E_{\vect{\theta}}[R(s, a, s^\prime)]$ is the gradient of the expectation. 
According to Sutton and Barto~\cite{sutton2018reinforcement}, $\nabla_{\vect{\theta}}E_{\vect{\theta}}[R(s, a, s^\prime)]$ is usually difficult to compute analytically, but can be efficiently approximated using rewards of actions sampled from the policy function $\pi_{\vect{\theta}}$ by: 
\begin{equation} 
\nabla_{{\vect{\theta}}}E_{\vect{\theta}}[R(s, a, s^\prime)]  \approx \frac{1}{|\widetilde{A}|} \sum_{a \in \widetilde{A}}
\nabla_{{\vect{\theta}}}\ln(\pi_{\vect{\theta}}(a|s)) \cdot R(s, a, s^\prime), 
\end{equation}
where $\widetilde{A}$ denotes the set of sampled actions. To reduce computational costs, Williams~\cite{williams1992simple} proposed a stochastic-gradient-ascent based algorithm (called REINFORCE in~\cite{sutton2018reinforcement}) to update the policy based on the reward of one single sampled action by: 
\begin{equation} \label{equa:REINFORCE}
\vect{\theta} = \vect{\theta} + \alpha \cdot \nabla_{\vect{\theta}}\ln(\pi_{\vect{\theta}}(a|s)) \cdot R(s, a, s^\prime). 
\end{equation}
The stochasticity is raised from the sampling procedure.

\textbf{RL for parallel particle tracing: } 
Each process has an agent responsible for assigning the blocks in the process to an appropriate process with a minimized local cost. At a specific execution time, each block movement decision is dependent on the current block assignment. 
This MDP of an agent can be described by a tuple $(S, A, \pi_{\vect{\theta}}, R)$, where the state of each agent of a process corresponds to which blocks are assigned to the process, and agents can take action to move blocks among processes and change the states. 

\textit{State}:
The state space $S$ consists of all combinations of data blocks. 
In particular, the agent of the process indexed by $l$ has a time-dependent state $s_{t,l}$ indicating which blocks are owned by process $l$ at time $t$. 
The state $s_{t,l}$ is corresponding to the block assignment at time $t$, i.e. $s_{t,l}=B_{t, l}$. 

\textit{Action}:
The action space $A$ consists of all actions for moving a block to a process. 
Specifically, the agent of a process $l$ has actions to move blocks out of $B_{t, l}$ to other processes and change the current state; note that keeping a block in process $l$ is a special yet valid action for the agent. 
Every agent considers the movement of one block at a time. The action $a_{i, l^\prime} \in A$ denotes moving block $i$, the block under the consideration, to process $l^\prime$. 

\textit{Policy function} $\pi_{\vect{\theta}}(\cdot)$:
Given the current state, the agent of process $l$ has a policy function (defined in Equation~\ref{equa:policy_function}), which returns probabilities of actions. 

\textit{Reward function} $R(\cdot)$:
The reward function, 
\begin{equation} \label{equa:reward_function}
\begin{aligned}
&R(s_{t-\Delta t,l}, a_{i, l^\prime}, s_{t,l}) \\ 
= \; & \texttt{local_exec_cost}(B_{t-\Delta t, l}, B_{t-\Delta t, l})  \\ - &  \texttt{local_exec_cost}(B_{t-\Delta t, l}, B_{t, l}),  
\end{aligned}
\end{equation}
is defined by the reduction of a local execution cost function (Equation~\ref{equa:local_cost_function}), after a block movement action $a_{i, l^\prime}$ has been taken. Here, $B_{t-\Delta t, l} \in \mathcal{B}_{t-\Delta t}$ with state $s_{t-\Delta t,l}$ represents the block assignment  before an action $a_{i, l^\prime}$ is taken, and $B_{t, l} \in \mathcal{B}_{t}$ with a new state $s_{t, l}$ denotes the assignment after block movement has occurred at time $t$. $\Delta t$ can be considered as the overhead of taking actions. 

\subsection{Design of Cost Functions} \label{sect:cost_functions}

Each process assesses the execution time imbalance locally by monitoring costs of friend processes, where processes $l$ and ${l^\prime}$ are friends if and only if the binary numbers of the ranks $l$ and ${l^\prime}$ only have one-bit difference following the lifeline technique~\cite{saraswat2011lifeline, binyahib2019lifeline}. 
Each process's local execution cost includes (1) the maximal execution cost and (2) the standard deviation of the execution costs among friend processes. Hence, the decrease of a local execution cost corresponds to the reduction of the local maximal execution time and the imbalance among friend processes, and is used for the reward computation. We define the local execution cost: 
\begin{equation} \label{equa:local_cost_function}
\begin{aligned}
\texttt{local_exec_cost}(B_{t-\Delta t, l}, B_{t, l}) \\
=  \max_{{l^\prime} \in \mathcal{N}_B(l)} \texttt{cost}(B_{t-\Delta t, l^\prime}, B_{t, l^\prime}) + \sigma_{t, l}, 
\end{aligned}
\end{equation}
where $\mathcal{N}_B(l)$ represents the set of friend processes of process $l$. The standard deviation $\sigma_{t, l}$ is
\begin{equation} 
\begin{aligned}
\sigma_{t, l} 
= \sqrt{\frac{\sum_{{l^\prime} \in \mathcal{N}_B({l})} (\texttt{cost}(B_{t-\Delta t, l^\prime}, B_{t, l^\prime}) - \mu_{t, l})^2}{|\mathcal{N}_B({l})|}}, 
\end{aligned}
\end{equation}
and the local average cost $\mu_{t, l}$ is
\begin{equation}
\begin{aligned}
\mu_{t, l} = \frac{\sum_{{l^\prime} \in \mathcal{N}_B({l})} \texttt{cost}(B_{t-\Delta t, l^\prime}, B_{t, l^\prime})}{|\mathcal{N}_B({l})|}, 
\end{aligned}
\end{equation}
Cost function $\texttt{cost}(\cdot)$ is designed to match the optimization requirements. 
The total cost for the block adjustment from $B_{t-\Delta t, l}$ to $B_{t, l}$ is estimated by: 
\begin{equation} \label{equa:cost_function}
\begin{aligned}
& \texttt{cost}(B_{t-\Delta t, l}, B_{t, l})  \\ 
= & \texttt{cost}_{a}(B_{t, l}) \\
 + & \texttt{cost}_{b}(B_{t-\Delta t, l}, B_{t, l}) \\ 
 +  & \texttt{cost}_{p}(B_{t-\Delta t, l}, B_{t, l}), 
\end{aligned}
\end{equation}
where $\texttt{cost}_{a}(B_{t, l})$ measures the estimated particle advection time of process $l$,  $\texttt{cost}_{b}(B_{t-\Delta t, l}, B_{t, l})$ measures the communication cost of block transfer from $B_{t-\Delta t, l}$ to $B_{t, l}$, and $\texttt{cost}_{p}(B_{t-\Delta t, l}, B_{t, l})$ measures the communication cost of transferring particles between the block sets when necessary. The three costs are measured by seconds and, hence, can be directly added together to form the total cost. In the following, we discuss the three cost functions in detail. 


\textit{Cost for particle advection computation:}
We estimate processes' particle advection time based on the workload estimation model's outcomes in~Section~\ref{sect:workload_estimation}. 
The workload estimation model gives an estimated particle advection time of each block. The cost $\texttt{cost}_{a}(B_{t, l})$ of a process $l$ is the sum of workloads of the blocks in the process, and is defined by:  
\begin{equation}
\texttt{cost}_{a}(B_{t, l}) = \sum_{i \in B_{t, l}} w_i, 
\end{equation}
where $w_i$ denotes data block $i$'s workload estimated by Equation~\ref{equa:workload_estimation}. 



\textit{Cost for block transfer communication:}
The movement of a data block from the current process to the other one requires additional communication costs. 
The cost of exchanging blocks to transform $B_{t-\Delta t, l}$ into $B_{t, l}$ is: 
\begin{equation}
\begin{aligned}
&\texttt{cost}_{b}(B_{t-\Delta t, l}, B_{t, l}) \\ 
= & |B_{t-\Delta t, l} \setminus (B_{t-\Delta t, l} \cap B_{t, l})| \cdot d_b^{\textrm{send}} \\
+ & |B_{t, l} \setminus (B_{t-\Delta t, l} \cap B_{t, l})| \cdot d_b^{\textrm{recv}}, 
\end{aligned}
\end{equation}
where ``$\setminus$'' is the set subtraction/difference operation, and $d_b^{\textrm{send/recv}}$ is the data send/receive cost per block given in~Section~\ref{sect:communication_cost}. 

\textit{Cost for particle transfer communication:} 
The particle transfer procedure consists of particle send and receive. 
Given a process $l$, it sends particles to other processes when the particles in data block $i$ owned by process $l$, for example, are now advected out of the block boundary and enter a neighboring block $j$ but block $j$ is not owned by process $l$ after the block adjustment. 
Similarly, process $l$ receives particles from other processes when particles in data block $j$ in other processes exit the block boundaries and enter block $i$ that is in process $l$ after the block adjustment. 
Hence, the total transfer cost of the process $l$ is: 
\begin{equation}
\begin{aligned}
& \texttt{cost}_{p}(B_{t-\Delta t, l}, B_{t, l}) \\
= & \sum_{i \in B_{t-\Delta t, l}}
\;\;\; \sum_{j \in \mathcal{N}_b(i) \; \textrm{and} \; j \notin B_{t, l}}  \widetilde{n}{(j|i)} \cdot d_p^{\textrm{send}}
\\
\\
+ & 
\sum_{i \in B_{t, l}}
\;\;\; \sum_{j \in \mathcal{N}_b(i) \; \textrm{and} \; j \notin B_{t-\Delta t, l}}  \widetilde{n}{(i|j)} \cdot d_p^{\textrm{recv}}, 
\end{aligned}
\end{equation}
where $\mathcal{N}_b(i)$ denotes the set of neighboring blocks of the given data block $i$, $\widetilde{n}{(j|i)}$ is the number of particles transferred from block $i$ to block $j$, and $d_p^{\textrm{send/recv}}$ is the data send/receive cost per particle given in~Section~\ref{sect:communication_cost}.

\subsection{Design of Policy Function and Update of Parameter}

We explain how to parameterize the policy function, which considers both workloads of processes and possible transfer costs by taking a state-action feature vector as an input for the decision-making of block movements. 
We then update the parameter of the policy function based on rewards. 

\textbf{Policy function:}
The policy function projects a latent state $z_{a_{i, l^\prime}}$ of an action $a_{i, l^\prime}$ to a probability. We parameterize the policy function, 
\begin{equation} \label{equa:policy_function}
\pi_{\vect{\theta}}(a_{i, l^\prime}|s_{t-\Delta t, l}) = \frac{e^{z_{a_{i, l^\prime}}}}{\sum_{a_{i, l^{\prime\prime}} \in A} e^{z_{a_{i, l^{\prime\prime}}}}}, 
\end{equation}
by parameter $\vect{\theta}$ using \texttt{Softmax} policy parameterization framework, which is shown to converge fast~\cite{agarwal2019theory, agarwal2020optimality, mei2020global}. 

\textbf{Latent state and function: }
A latent state of an action is a scalar and represents the value of the action, where the scalar is larger indicating the action is more favored. A latent state of an action $a_{i, l^\prime}$ is mapped from an observed feature vector $\vect{\phi}(s_{t-\Delta t, l}, a_{i, l^\prime})$ using a latent function, 
\begin{equation} \label{equa:latent_function}
z_{a_{i, l^\prime}} = f_{\vect{\theta}}(\vect{\phi}(s_{t-\Delta t, l}, a_{i, l^\prime})) = \frac{1}{w_i} \vect{\phi}(s_{t-\Delta t, l}, a_{i, l^\prime}) \cdot \vect{\theta}, 
\end{equation}
where $\frac{1}{w_i}$, reciprocal of block $i$'s workload, is used for the normalization for making decision across different blocks, and the bias term is omitted for brevity.
Intuitively, the latent function with a linear form represents the weighted combinations of the feature vector components. The weights in $\vect{\theta}$ are non-negative and indicate different features' importances. 
The linear latent function can be computed with low overheads and is usually applied in studies of decision-making~\cite{glascher2010states, reverdy2015parameter}. The linear latent function combined with the \texttt{Softmax} policy is categorized into the log-linear policy class in RL~\cite{agarwal2019theory}. 


\textbf{State-action feature vector:}
A feature vector $\vect{\phi}(s_{t-\Delta t, l}, a_{i, l^\prime})$ is a three-dimensional vector formed to describe the observed information, including the improvements of workload imbalance and data transfer cost associated with the action to move block $i$ to process $l^\prime$. The feature vector consists of three components. 

The first component is to encourage to move block $i$ from process $l$ to the other process $l^\prime$ with a lower workload, and is computed by the workload of process $l$ (excluding the workload of block $i$) reduced by the workload of $l^\prime$: 
\begin{equation}
\begin{aligned}
& \vect{\phi}_1(s_{t-\Delta t, l}, a_{i, l^\prime}) \\
= & \texttt{cost}_{a}(B_{t-\Delta t, l} \setminus \{i\}) - \texttt{cost}_{a}(B_{t-\Delta t, l^\prime} \setminus \{i\}). 
\end{aligned}
\end{equation}
If the workload of $l$ excluding block $i$ is higher than the workload of $l^\prime$, this feature component becomes positive, encouraging the action. Note that this feature component is zero if $l^\prime$ equals $l$, i.e., if the action is to make block $i$ stay in process $l$. 

The second and third components correspond to the decrements of block transfer cost and particle transfer cost, respectively, when moving a block $i$ from process $l$ to $l^\prime$. 
The second component is: 
\begin{equation}
\vect{\phi}_2(s_{t-\Delta t,l}, a_{i, l^\prime}) =
\left\{
\begin{aligned}
&0, &&\textrm{if}\ l=l^\prime \\
&-(d_b^{\textrm{send}} + d_b^{\textrm{recv}}),  &&\textrm{otherwise}. 
\end{aligned}
\right.
\end{equation}
The third component is: 
\begin{equation}
\begin{aligned}
& \vect{\phi}_3(s_{t-\Delta t, l}, a_{i, l^\prime}) \\
= & \sum_{j \in \mathcal{N}_b(i) \; and \; j \in B_{t-\Delta t, l^\prime}}  \widetilde{n}{(i|j)} \cdot (d_p^{\textrm{send}} + d_p^{\textrm{recv}})
\\ 
- & \sum_{j \in \mathcal{N}_b(i) \; and \; j \in B_{t-\Delta t, l}} \widetilde{n}{(i|j)} \cdot (d_p^{\textrm{send}} + d_p^{\textrm{recv}}). 
\end{aligned}
\end{equation}
Transfers of the particles that enter block $i$ get changed after block $i$ is moved from process $l$ to process $l^\prime$. The part of the particles owned by process $l^\prime$ will not need to be transferred, where the transfer cost is saved. However, the part of the particles owned by process $l$ will need to be transferred to process $l^\prime$ after the block movement.

\textbf{Policy parameter update:}
The parameter $\vect{\theta}$ is updated to maximize the expected rewards of sampled actions, encouraging actions to approach the minimum of the local load cost function. 
Following Equation~\ref{equa:REINFORCE}, we update the policy parameter $\vect{\theta}$ using stochastic gradient ascent by: 
\begin{equation} \label{equa:gradient_ascent}
\vect{\theta} = \vect{\theta} + \alpha \cdot \nabla_{\theta}\ln(\pi_{\vect{\theta}}(a_{i, l^\prime}|s_{t-\Delta t, l})) \cdot R(s_{t-\Delta t, l}, a_{i, l^\prime}, s_{t, l}). 
\end{equation}

\section{High-Order Workload Estimation Model} \label{sect:workload_estimation}

We estimate the particle advection workload in a given block using a high-order estimation model for advection cost computations when RL agents move blocks. Given a data block, the input of this workload estimation model consists of the statistics of the incoming particles to the data block; the model's output is the expected advection workload of those particles. 

The blockwise workload estimation is based on the historical records of particles that have been advected previously. When a particle enters a data block, the total number of advection steps is deterministic, although unknown before the advection occurs. For a given data block, the existing zeroth-order model~\cite{peterka2011study, zhang2018dynamic} assumes every entering particle has a similar number of advection integration steps. Hence, the zeroth-order model averages numbers of integration steps of particles that have been advected in the given block for the estimation of incoming particles. 

Our high-order model utilizes the closeness of particle trajectories. By assuming continuities exist in spacetime domain following~\cite{tricoche2002topology, garth2004tracking}, particles with close-by entry points into a data block usually have similar numbers of advection steps; also, particles whose trajectories are close to one another in the domain before entering the block tend to have similar entry points. Because matching the full trajectories of particles leads to high overhead, we record only the sequence of blocks accessed by those particles as an approximation of the trajectories, at a much lower cost. 
This is inspired by the particle access dependency modeling of Zhang et al. \cite{zhang2016efficient}, which discretizes the trajectories of particles into sequences of data blocks accessed by the particles and then uses a high-order Markov chain~\cite{raftery1985model} to estimate the particles' data access patterns. 

An additional parameter $r$ is used to control how many data blocks most recently accessed by each particle are to be recorded, where $r$ is the \textit{order} of our estimation model and controls the model complexity. Existing works~\cite{peterka2011study, zhang2018dynamic} use the historical average of the number of advection steps to estimate the workload of incoming particles, which is a special case when $r$ is zero (zeroth-order) in our model.

\begin{figure}[htb]
\centering

  \includegraphics[width=1.0\columnwidth]{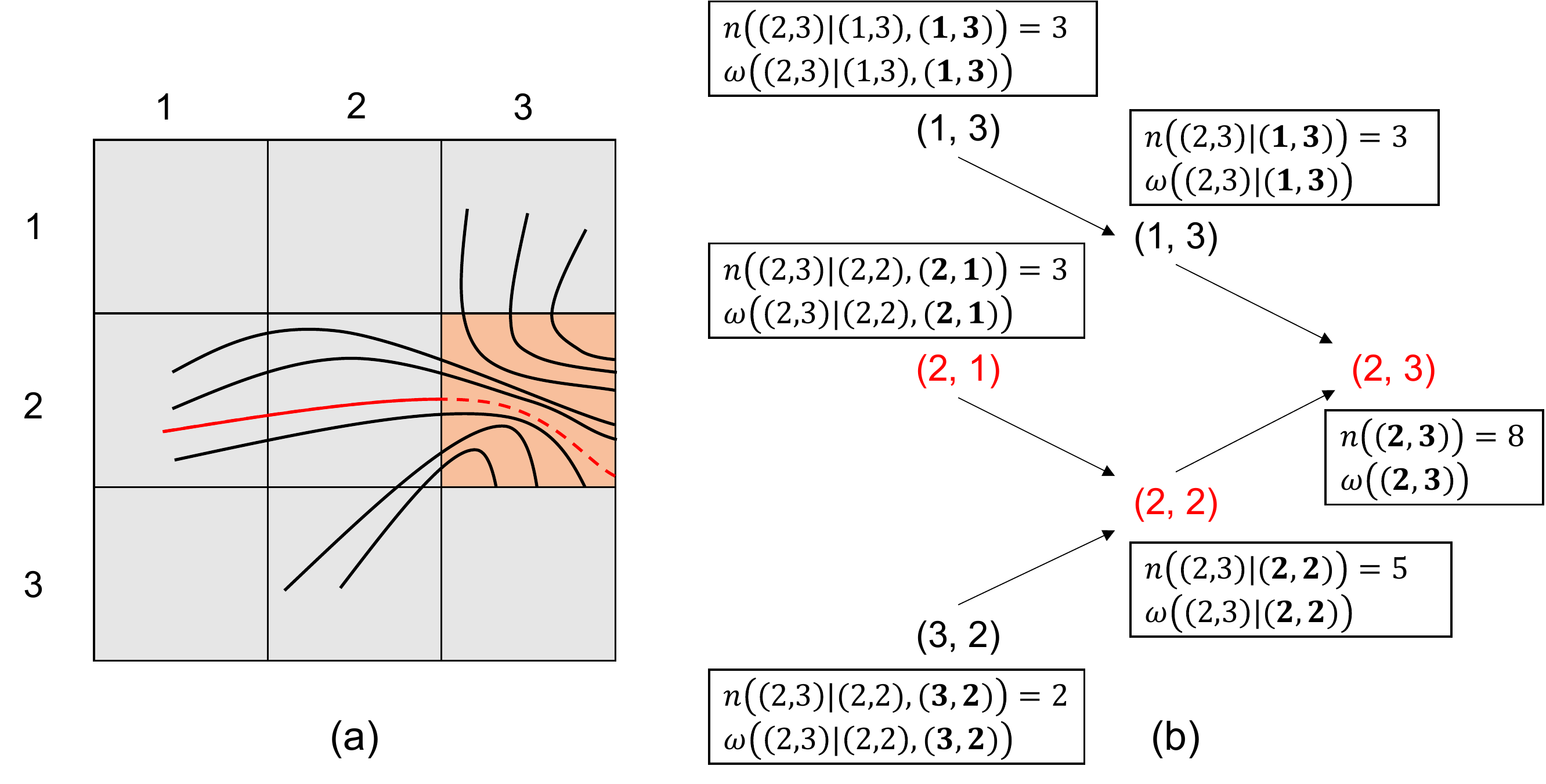}

  \caption{An illustration of high-order advection workload estimation. (a) A 2D example. Eight particles, whose trajectories are colored in black, have been traced previously within the block $(2, 3)$ with actual numbers of advection steps recorded. A newly-incoming particle is colored in red. 
  The zeroth-order estimation model uses the average of the numbers of advection steps of the eight particles traced to estimate the workload of the incoming particle. While, the high-order estimation model uses the numbers of the advection steps of the three particles close to the incoming particle and also passed through block $(2, 1)$, $(2, 2)$, and $(2, 3)$ for the estimation. 
  (b) The corresponding trajectories tree. We represent the high-order (second-order in this example) workload estimation model by abstracting the particles' accessed blocks using a tree structure rooted at $(2, 3)$ with a depth of two. 
  }

  \label{fig:workload_estimation_diagram}
\end{figure}

Our method has two steps, as illustrated in~Fig.~\ref{fig:workload_estimation_diagram}. First, we match the trajectories of incoming particles with those of the previously advected particles. 
Second, our method uses the historical particles' numbers of advection steps that have the best-matched trajectories with the particles in question to estimate the workload of advection in the current block.

\subsection{Data Structure of Trajectories Tree}

Our model's data structure is a tree, called trajectories tree, which groups particles based on where they come from. Trajectories trees are created from the block sequences accessed by the previously advected particles; the depth of the tree is $r$. Every block has a corresponding tree structure. 
For example, in Fig.~\ref{fig:workload_estimation_diagram}b, the tree is formed by the access sequences of blocks by the eight black-colored particles in Fig.~\ref{fig:workload_estimation_diagram}a and is stored in the process that owns block $(2, 3)$. 


Each tree node has two attributes: $n(\cdot)$ and $\omega(\cdot)$, which are historical statistics computed from the particles that access data blocks following the path from the tree node to the root. 
$n({i_{r}|i_{r-1},...,i_{0}})$ denotes the number of previoulsy advected particles that travel through the blocks of $i_0$, ..., $i_{r-1}$  to $i_r$. $\omega({i_{r}|i_{r-1},...,i_{0}})$ represents the historical average of the numbers of advection steps of particles that travel from blocks $i_{0}$ to $i_{r}$. This tree data structure is saved in the process owning the root block $i_{r}$. 


\subsection{Workload Estimation of Advection Computation}
We estimate the time of particle advection in data block  $i_r$ based on the incoming particles' trajectories. 
For example, for the new particle (red particle in Fig.~\ref{fig:workload_estimation_diagram}a) that is going to access block $(2, 3)$, we search the trajectories tree in Fig.~\ref{fig:workload_estimation_diagram}b top-down to find a match of the particle's block based trajectory with those of the previously advected particles for workload estimation. 
Because the incoming particle traveled through blocks $(2,1)$, $(2,2)$, and $(2, 3)$ in order, we use the attribute $\omega((2,3)|(2,2),(2,1))$ stored in node $(2,1)$ to estimate the workload of the particle; $\omega((2,3)|(2,2),(2,1))$ is the average advection step of the black-colored three particles starting from block $(2,1)$. 

\textbf{Estimation:} 
More formally, the advection time, noted by $w_{i_r}$, of all incoming particles to the block $i_r$ is estimated by: 
\begin{equation} \label{equa:workload_estimation}
\begin{aligned}
    w_{i_r} 
    = 
    \sum_{{i_{r-1}} \in \mathcal{N}_b(i_{r})}...\sum_{{i_{0}} \in \mathcal{N}_b({i_{1}})}
    & \widetilde{n}({i_{r}|i_{r-1},...,i_{0}}) \cdot d_a
    \\
    &  \cdot \omega({i_{r}|i_{r-1},...,i_{0}}), 
\end{aligned}
\end{equation}
where $\mathcal{N}_b(\cdot)$ gives neighboring data blocks, $d_a$ is the historical time cost per advection step, and $\widetilde{n}({i_{r}|i_{r-1},...,i_{0}})$ denotes the number of the particles that have their block-based trajectories as $i_0$, ..., $i_{r-1}$ and are now going to enter the block $i_{r}$. $\widetilde{n}(\cdot)$ is computed and exchanged among the parallel processes before the workload estimation. 

\textbf{Boundary case:} 
If a trajectories tree data structure has no records of $\omega({i_{r}|i_{r-1},...,i_{0}})$, we will use an existing $\omega({i_{r}|i_{r-1},...,i_{k}})$ with the smallest $k$ to approximate $\omega({i_{r}|i_{r-1},...,i_{0}})$. For example, in Fig.~\ref{fig:workload_estimation_diagram}, if a new particle accesses blocks $(1,2)$, $(1,3)$, and $(2,3)$, we will use the attributes stored on tree node $(1,3)$ to estimate the workload. 

\subsection{Online Update of Estimation Model}

We update the attributes, $n(i_{r}|\cdot)$ and $\omega(i_{r}|\cdot)$, stored in the trajectories tree nodes every time after particles finish advection within block $i_r$. 

\textbf{Update of $\boldsymbol{n(i_{r}|\cdot)}$:} 
$n({i_{r}|i_{r-1},...,i_{0}})$ is updated directly based on the blocks that the advected particles have passed through; if particles have accessed less than $r$ blocks, we use their seeding block ID to pad the trajectory of the block sequence until the length becomes $r$. Then, we update the aggregate statistics $n({i_{r}|i_{r-1},...,i_{k}})$, which indicates the sum of all particles that appear in the block $i_{r}$ and previously pass through the blocks from $i_k$ to $i_{r-1}$, using: 
\begin{equation}
    n({i_{r}|i_{r-1},...,i_{k}}) =  \sum_{{i_{k-1}} \in \mathcal{N}_b(i_{k})} n({i_{r}|i_{r-1},...,i_{k},i_{k-1}}), 
\end{equation}
where $k$ is looped starting from $1$ to $r-1$ for the update from the bottom to the top of the trajectories tree. Finally, we update the tree root: 
\begin{equation}
    n({i_{r}}) =  \sum_{{i_{r-1}} \in \mathcal{N}_b(i_{r})} n({i_{r}|i_{r-1}}). 
\end{equation}

\textbf{Update of $\boldsymbol{\omega({i_{r}|\cdot})}$:} 
$\omega({i_{r}|i_{r-1},...,i_{0}})$ is updated directly after each particle finishes its advection within data block $i_{r}$; if the length of a particle's block-based trajectory sequence is less than $r$, we pad the block sequence using its initial seeding block ID until the sequence length becomes $r$. We update $\omega({i_{r}|i_{r-1},...,i_{k}})$ by averaging the numbers of advection steps of those particles in block $i_{r}$ and previously pass through blocks from $i_k$ to $i_{r-1}$ using $\omega({i_{r}|i_{r-1},...,i_{k},i_{k-1}}), \forall {i_{k-1}} \in \mathcal{N}_b(i_{k})$: 
\begin{equation}
\begin{aligned}
    &\omega({i_{r}|i_{r-1},...,i_{k}})
    \\
    = & \frac{\sum_{{i_{k-1}} \in \mathcal{N}_b(i_{k})} n({i_{r}|i_{r-1},...,i_{k-1}}) \cdot \omega({i_{r}|i_{r-1},...,i_{k-1}})}{n({i_{r}|i_{r-1},...,i_{k}})}, 
\end{aligned}
\end{equation}
where $k$ is looped starting from $1$ to $r-1$ for the bottom-up update of the tree. Finally, we update the tree root: 
\begin{equation} 
    \omega({i_{r}}) = \frac{\sum_{{i_{r-1}} \in \mathcal{N}_b(i_{r})} n({i_{r}|i_{r-1}}) \cdot \omega({i_{r}|i_{r-1}})}{n({i_r})}. 
\end{equation}

\section{Communication Cost Model}
\label{sect:communication_cost}

We build linear transmission cost models~\cite{kielmann2001network, chan2007collective, traff2019optimal} to estimate communication costs of data transfers. The linear cost model considers that the time cost of transferring a data entity from one process to the other is proportional to the size of the entity plus a fixed starting latency. Because the sizes of data blocks are similar and the sizes of particles are the same, we assume moving each block has a similar cost, and so does moving each particle. 
Linear models can be fitted efficiently, hence, causing low overheads. 

\textbf{Historical-record collection:}
We collect the records of four types of data transfer events: data block (including trajectories tree) send/receive and particle send/receive. Generally, each record is a tuple, where the first item is the number of involved entities (noted as $x$) in an event, and the second item is the total time (noted as $y$) of an event.






\textbf{Model fitting:}
For each event, we build a linear model
\begin{equation}
y = d \cdot x + e, 
\end{equation}
where $e$ is a constant latency of the event, and $d$ is an additional time cost for each additional entity. The least-squares method is used to fit the linear model from historical records of the event. 
After the fitting, we obtain the send/receive time cost per data block $d_b^{\textrm{send/recv}}$ and the send/receive time cost per particle $d_p^{\textrm{send/recv}}$.

\begin{table*}[tb]
  \caption{Specifications of four datasets. We seed uniformly on Nek5000 and Turbulence to trace streamlines, seed locally on Ocean data to produce pathlines, and seed at all grid points of Isabel data to generate a FTLE field. }
  \label{tab:datasets}
  \scriptsize%
	\centering%
  \begin{tabu}{%
	*{8}{c}%
	}
  \toprule
   Dataset & Domain & Timestep & Size & Visualization Application
  & Seed Count & Maximum Advection Steps   \\
  \midrule
    Nek5000 & $512 \times 512 \times 512$ & $1$ & $1.5$ GiB & Streamlines
    & $2$M & $1,024$  \\
    
    Ocean & $3,600 \times 2,400 \times 1$  & $36$ & $2.3$ GiB & Pathlines 
    & $2.7$M & $1,024$  \\
    
    Isabel & $500 \times 500 \times 100$ & $48$ & $13.4$ GiB & FTLE  
    & $25$M & $48$  \\
    Turbulence & $4,096 \times 4,096 \times 4,096$ & $1$ & $768$ GiB & Streamlines
    & $2.6$M, $16.8$M, $134.2$M & $1,024$  \\
  \bottomrule
  \end{tabu}%
\end{table*}

\begin{figure*}[tb]
\centering
\captionsetup[subfigure]{labelformat=empty}

 \begin{subfigure}[b]{0.194\linewidth} 
   \includegraphics[width=\linewidth]{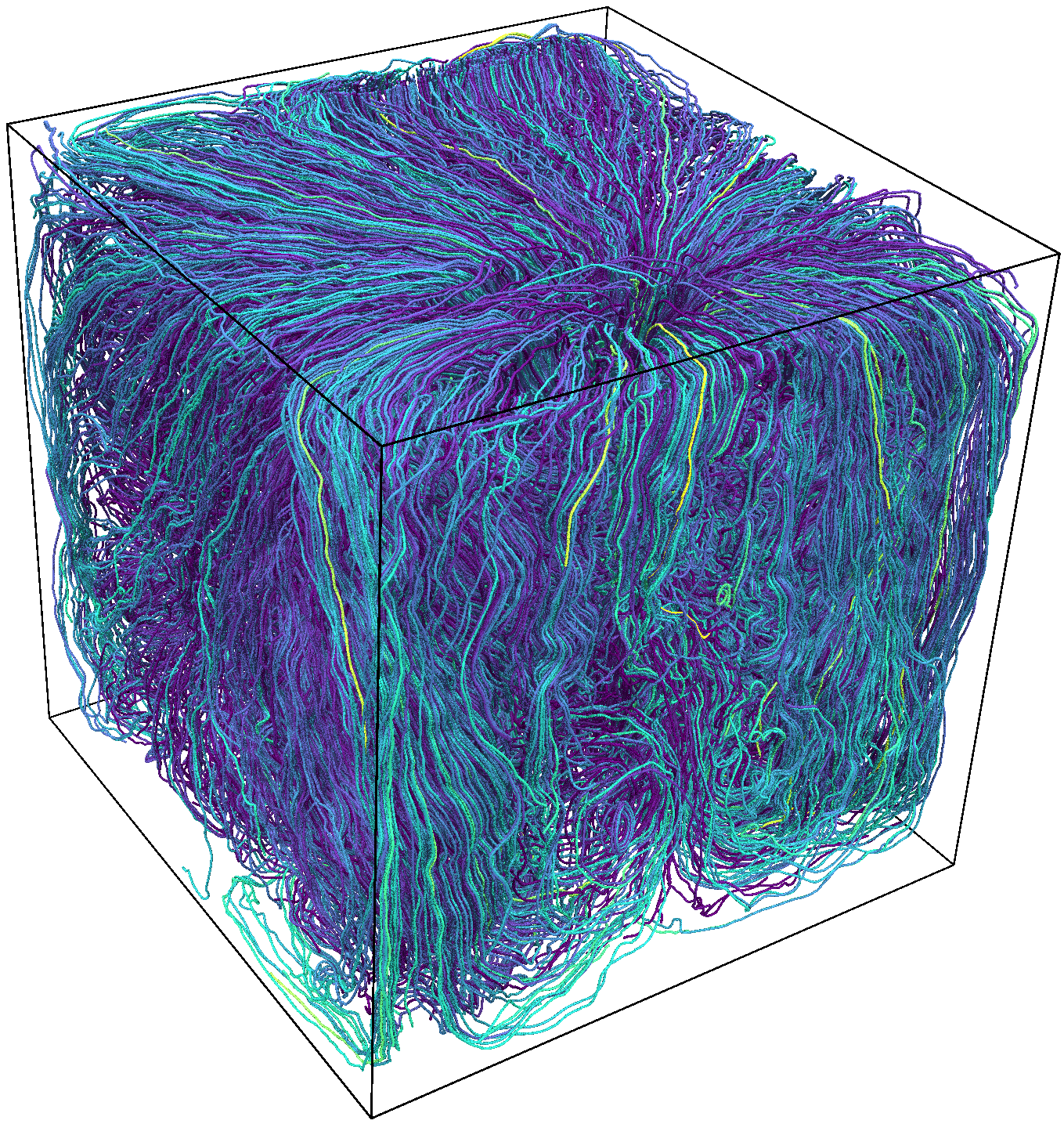}
   \caption{(a) Nek5000} \label{fig:streamline}
 \end{subfigure}
 \hfill
 \begin{subfigure}[b]{0.311\linewidth} 
   \includegraphics[width=\linewidth]{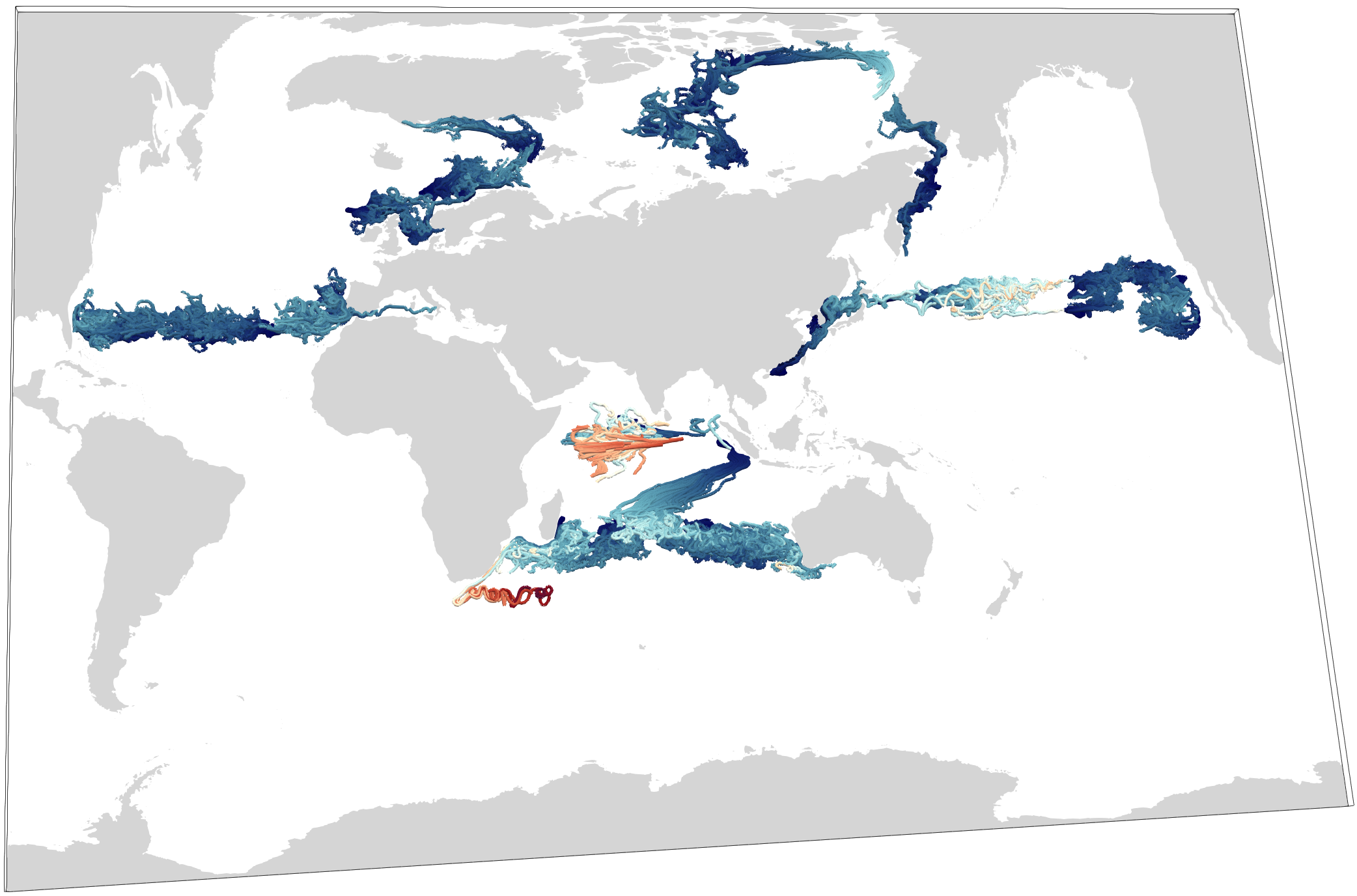}
   \caption{(b) Ocean} \label{fig:pathline}
 \end{subfigure} 
 \hfill
 \begin{subfigure}[b]{0.233\linewidth} 
   \includegraphics[width=\linewidth]{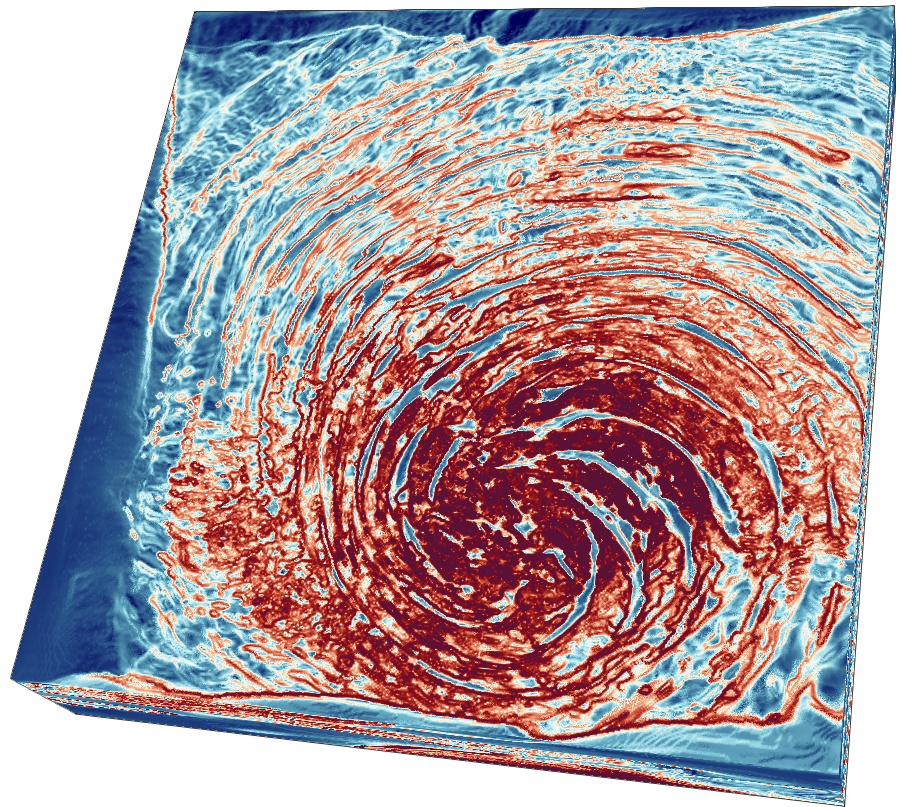}
   \caption{(c) Isabel} \label{fig:FTLE}
 \end{subfigure} 
\hfill
 \begin{subfigure}[b]{0.211\linewidth} 
   \includegraphics[width=\linewidth]{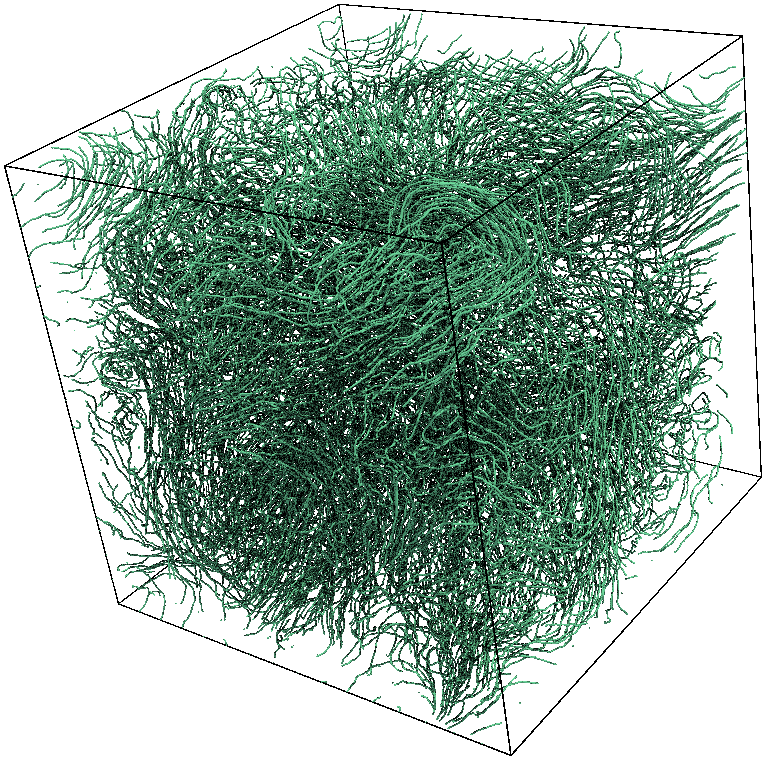}
   \caption{(d) Turbulence} \label{fig:streamline_isotropic}
 \end{subfigure} 

  \caption{Examples of rendering results. (a) We generated $4,096$ streamlines on Nek5000 dataset for static flow analysis. (b) We used $2,592$ pathlines for the Ocean dataset, which are seeded near Eurasia for the source-destination query. (c) We tested the Isabel dataset by using an FTLE field at timestep 0 within a time range of 16. (d) We generated $4,096$ streamlines using the Turbulence dataset for flow turbulence analysis.  }

  \label{fig:phases}
\end{figure*}

\section{Performance Evaluation}
\label{sect:rl_evaluation}

We studied our method's performance using four particle-tracing tasks, including static flow analysis on a Nek5000 dataset via streamlines, source-destination analysis on an Ocean dataset via pathlines, unsteady flow analysis on an Isabel dataset via FTLEs, and static turbulence analysis on a Turbulence dataset via streamlines.

\textbf{Datasets:}
Table~\ref{tab:datasets} describes the detailed specifications of the four datasets. The \textit{Nek5000} dataset is a thermal-hydraulics dataset produced by the Nek5000 solver of a large-eddy Navier-Stokes simulation \cite{fischer2008petascale}. We took one timestep of the simulation output to analyze the flow patterns of the fluid dynamics statically using streamlines (shown in Fig.~\ref{fig:streamline}). The \textit{Ocean} dataset was produced by an eddy-resolving simulation with $1 \backslash 10^{\circ}$ horizontal spacing \cite{maltrud2005eddy}. The dataset consists of monthly averaged time-varying vector fields from February 2001 to January 2004. 
Source-destination query analysis was performed with local dense seeding by using pathline visualizations (shown in Fig.~\ref{fig:pathline}), where a source-destination query is related to seed particles in local regions and queries the destination of those seeded particles. 
The hurricane \textit{Isabel} data was produced from an atmospheric simulation developed by the National Center for Atmospheric Research. We performed an unsteady flow analysis using FTLEs (shown in Fig.~\ref{fig:FTLE}). The isotropic \textit{Turbulence} dataset is a direct numerical simulation of turbulent fluid flow on a $4096^3$ grid~\cite{yeung2012dissipation} and is maintained by the Johns Hopkins Turbulence Databases~\cite{li2008public}.
When the simulation achieved a statistically stationary state, one snapshot of data was output and analyzed statically using streamlines (shown in Fig.~\ref{fig:streamline_isotropic}). 

We have different seeding settings for the analysis of the four datasets. For the static flow analysis on Nek5000 and Turbulence data, we uniformly seeded particles to generate streamlines. For the source-destination query on Ocean, we locally seeded particles near Eurasia, consisting of Europe and Asia. For the unsteady flow analysis Isabel, we generated an FTLE field by tracing from all grid points.

\textbf{Computing platforms specifications:} We evaluated Nek5000, Ocean, and Isabel data on Bebop high-performance computing (HPC) cluster and tested Turbulence data on Theta supercomputer. 

\textit{Bebop} HPC cluster has $664$ compute nodes and uses IBM General Parallel File System. Every compute node has $32$ Intel Xeon E5-2695v4 CPU cores with $4$ GB memory per core. The compute nodes are interconnected by an Intel Omni-Path network. Message passing uses the Intel MPI library. We used up to $1,024$ cores for the following studies. 

\textit{Theta} supercomputer has $4,392$ compute nodes and uses Lustre Parallel File System. Every compute node has $64$ Intel Xeon Phi 7230 processors and $3$ GiB memory per processor. The compute nodes are interconnected by Cray's Aries technology and integrated by the Dragonfly network topology. We used up to $16,384$ processors on Theta. 

\textbf{Implementation details:}
We prototyped our methods based on \texttt{Python 3}~\cite{van2009python}. 
The advection integral was implemented using \texttt{C}~\cite{ritchie1988c} programs for fast computation and then imported into \texttt{Python} with the support of \texttt{Cython}~\cite{behnel2010cython}. 

Processes' communications and I/O were supported by \texttt{mpi4py}~\cite{dalcin2005mpi} and \texttt{DIY}~\cite{peterka2011scalable, morozov2016block, morozov2021diy} libraries with asynchronous parallelism and nonblocking communications. 
\texttt{mpi4py} library~\cite{dalcin2005mpi} offered python-based Message-Passing Interface (MPI) implementations and was used for the communications amongst processes. The efficiency of the block-structured parallel I/O was improved by using the block-IO layer~\cite{kendall2011toward} supported by \texttt{DIY} library~\cite{peterka2011scalable, morozov2016block, morozov2021diy}. 
To store the block assignment $\mathcal{B}_t$ in memory, a block-to-process mapping recorded which blocks belong to which processes, and was maintained in distributed memory using \texttt{MPI} one-sided communications with low overhead, following the dynamic assignment algorithm of DIY library~\cite{peterka2011scalable, morozov2016block, morozov2021diy}. 

\texttt{PyTorch} library~\cite{paszke2019pytorch} supported the training of RL agents. 
\texttt{autograd}~\cite{paszke2017automatic} module supported automatic differentiation for gradient and derivative computation, which was then used to update the parameters of agents through backpropagation~\cite{rumelhart1985learning} and RMSProp~\cite{Tieleman2012} optimizer. 
RMSProp can adapt the learning rate (i.e., $\alpha$ in Equation~\ref{equa:gradient_ascent}) automatically and was reported~\cite{kingma2014adam} to handle non-stationary environments well. Hence, RMSProp is commonly used in reinforcement learning studies, such as ~\cite{mnih2013playing, mnih2015human, mnih2016asynchronous}. 

Eight blocks were assigned to each process initially. The block assignment was then adjusted dynamically by our work donation algorithm during the runtime.

\textbf{Baseline:}
We implemented the most recently published lifeline-based work stealing/requesting approach~\cite{binyahib2019lifeline}. The lifeline-based approach~\cite{binyahib2019lifeline} was shown to improve performance compared with previous work stealing/requesting methods~\cite{muller2013distributed, lu2014scalable} and thus is one of the current state-of-the-art in dynamic load balancing for parallel particle tracing.

\begin{figure*}[tb]
  \centering 
  \includegraphics[width=\linewidth]{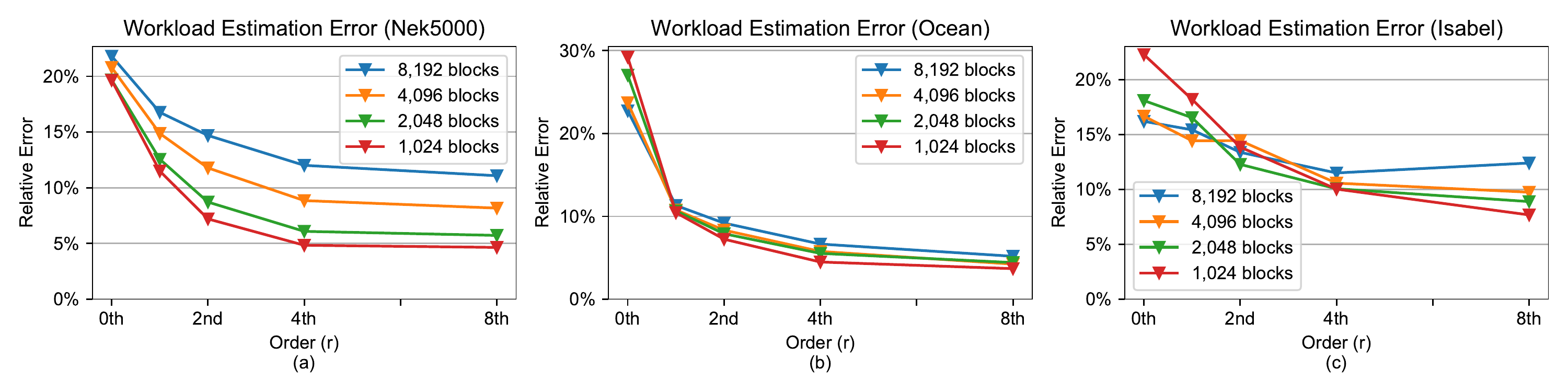}
  \caption{Advection workload estimation errors (defined in Section~\ref{sect:eval_workload_estimation}) under different processes and order settings. Three subfigures are corresponding to the three datasets: Nek5000, Ocean, and Isabel, respectively. }
  \label{fig:workload_estimation_errors}
\end{figure*}

\subsection{Advection Workload Estimation Study}
\label{sect:eval_workload_estimation}

We evaluated the accuracy of our high-order workload estimation model. 
Fig.~\ref{fig:workload_estimation_errors} shows the estimation errors when the workload estimation model uses varying orders, where the order represents the length of the recorded accessed blocks for each particle and indicates the complexity of the model. 
The \textit{relative error} is computed by the sum of the absolute difference between estimated advection time and actual advection time in each block divided by the entire advection time. 

The estimation error generally decreases as the order increases with different total block counts for domain decomposition, where higher block counts correspond to smaller block sizes. 
The result of zeroth-order has errors of around $20\%$ for Nek5000 data, $23\% \sim 29\%$ for Ocean data, and $16\% \sim 22\%$ for Isabel data. With the order becomes eighth, the errors become  $5\% \sim 11\%$ for Nek5000 data, $4\% \sim 5\%$ for Ocean data, and $8\% \sim 12\%$ for Isabel data. 

Errors are difficult to decrease further after model orders are high enough because of the turbulence in vector fields. Turbulence makes the vector field of a data block not continuous and causes that particles enter at similar entry positions into a data block yet still have different numbers of advection steps. 
We select fourth-order as the order setting for our following performance study because the errors do not decrease much when using orders higher than four. 

\subsection{Performance Study}

To evaluate the performance of our method, we conducted the evaluation by using three measurements: 
\begin{enumerate} 

\item \textbf{Strong scaling:} The strong scaling measures the efficiency of an approach on a fixed-size problem with different process counts. By fixing the number of particles, this measurement evaluates how the execution time changes along with increasing processes. Optimal strong scaling is achieved when the execution time is inversely proportional to the number of processes. 

\item \textbf{Advection load imbalance:} The imbalance is measured by the metric ``$\frac{\textrm{MAX}}{\textrm{AVG}}$'', which is the maximal particle tracing time over all processes divided by the average tracing time per process, following previous studies~\cite{zhang2017dynamic, zhang2018dynamic}. When the advection workload of processes is highly imbalanced, the value of this metric is large. As the workload of processes becomes similar, the metric is approaching $1$. 

\item \textbf{I/O and communication cost:}
The I/O and communication cost measures the average time per process used for both data blocks' loading and interprocess exchange of data blocks and particles. We combine them together because both I/O and communication time are related to the data blocks' fetching during the runtime. 

\end{enumerate}

\begin{figure}[tb]
  \centering 
  \includegraphics[width=\linewidth]{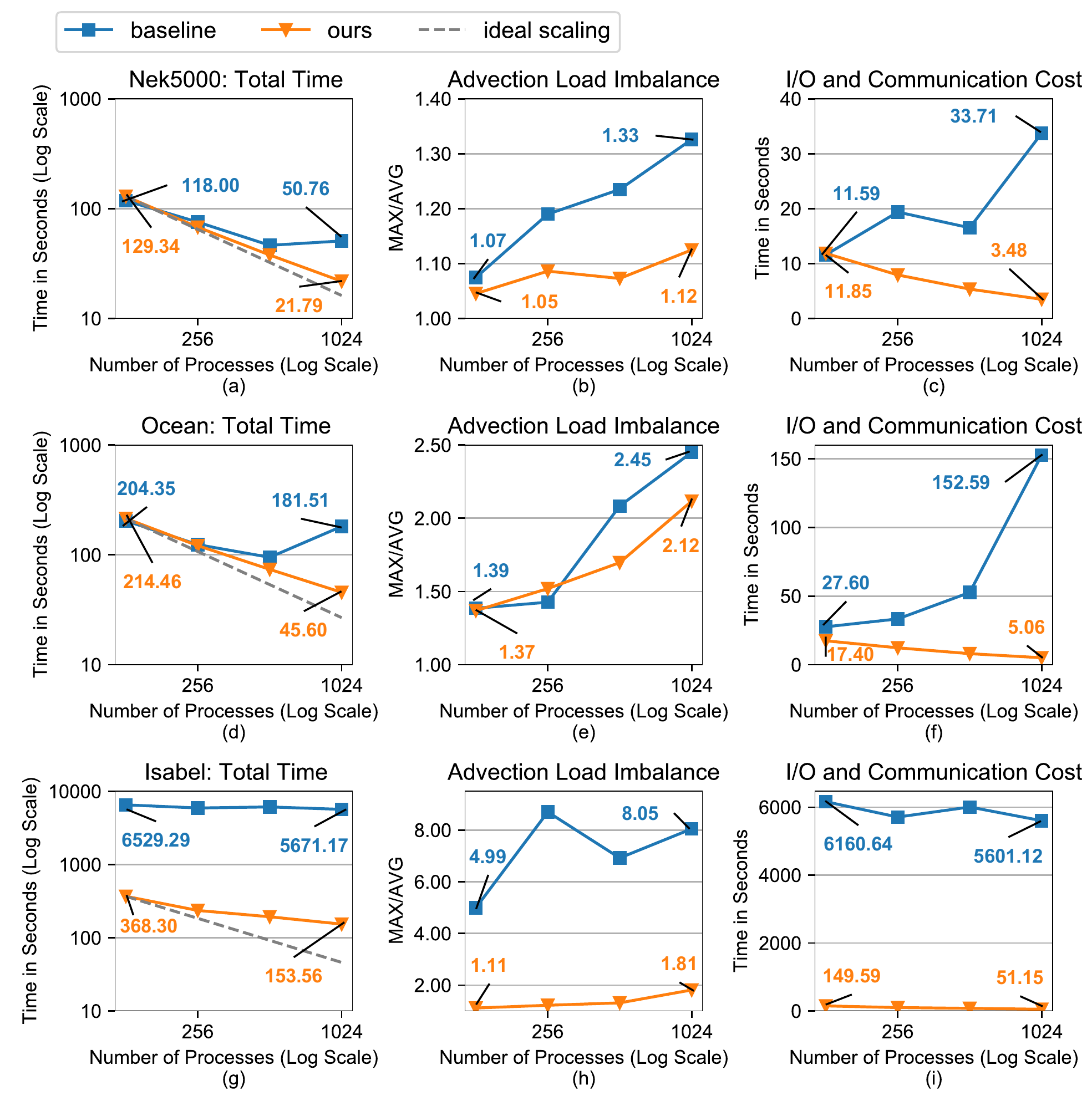}
  \caption{Performance comparisons between the baseline and our method on the Bebop HPC cluster. 
  Three rows correspond to the studies on three datasets, including Nek5000, Ocean, and Isabel. 
  The results of three measurements, including total execution time, load imbalance, and I/O and communication time, are presented in the three columns. }
  \label{fig:strong_scaling_platforms}
\end{figure}

\textbf{Strong scalability study:}
The first column of Fig.~\ref{fig:strong_scaling_platforms} displays the total execution time to evaluate our method's strong scaling performance up to using $1,024$ processes. As shown in Fig.~\ref{fig:strong_scaling_platforms} (a), (d), and (g), our approach had a lower total execution time than the baseline across the three test datasets when $1,024$ processes are used. Also, our method's total running time decreased faster than the baseline did as the process count grew. 
Our method's overheads were $1.35$, $0.92$, and $12.33$ seconds for the three datasets when $1,024$ processes were used, which were $6.19\%$, $2.02\%$, and $8.03\%$ of the total execution time. 
Also, compared with the ideal scaling, the baseline attained the strong scaling efficiency of $29.06\%$, $14.07\%$, and $14.39\%$ for the three datasets. 
Our method attained the parallel efficiency of $74.19\%$, $58.79\%$, and $29.98\%$ for the three datasets, respectively, with a speedup of $2.33$x, $3.98$x, and $36.93$x over the baseline. 
The speedup came from the improvement in both the workload imbalance and the costs of I/O and communication, which are demonstrated below.

\textbf{Advection workload imbalance study:}
The second column of Fig.~\ref{fig:strong_scaling_platforms} displays the advection workload imbalance for our method and the baseline approach. Lower values are better and indicate the ratio of the maximal advection workload to the average workload per process is smaller. 
The advection workload imbalance is important for parallel programs to minimize the idle time of processes, improving the total parallel efficiency. The workload imbalance for the three datasets evaluates whether our algorithm is general enough to dynamically optimize the workloads of processes for different tasks. 

As shown in Fig.~\ref{fig:strong_scaling_platforms} (b), (e), and (h), our method achieved lower imbalances than the baseline approach across the three datasets, although imbalance for both methods slightly increased as the process count grew. 
When $1,024$ processes were used, the baseline achieved the load imbalance of $1.33$, $2.45$, and $8.05$ for the three test datasets, respectively. 
Correspondingly, our method achieved the load imbalance of $1.12$, $2.12$, and $1.81$. 

The results demonstrate proactively balancing workloads before processes become idle was effective. 
RL agents dynamically donated work from overloaded processes to underloaded processes, hence, explicitly reducing the maximal workload over processes and improving the workload imbalance $\frac{\textrm{MAX}}{\textrm{AVG}}$ with minimal overheads. 

\textbf{I/O and communication costs study:}
The third column of Fig.~\ref{fig:strong_scaling_platforms} shows the I/O cost used for data block loading and the communication cost for the processes' exchange of both data blocks and particles. 
As shown in Fig.~\ref{fig:strong_scaling_platforms} (c), (f), and (i), our method outperformed the baseline across the three datasets. 
When $1,024$ processes were used, the baseline had the cost of $33.71$, $152.59$, and $5601.12$ seconds for the three test datasets, respectively. 
Our method had the cost of $3.48$, $5.06$, and $51.15$ seconds, which are $10.33\%$, $3.32\%$, and $0.91\%$ of the time spent by the baseline. 
Compared with the baseline~\cite{binyahib2019lifeline} that chooses to load data blocks from disks, our method is designed to fetch data blocks from other processes through network transfer and minimizes the communication cost for data exchange when optimizing the total execution time, which reduced the I/O and communication time across the three applications.

\begin{figure*}[bt]
  \centering 
  \includegraphics[width=\linewidth]{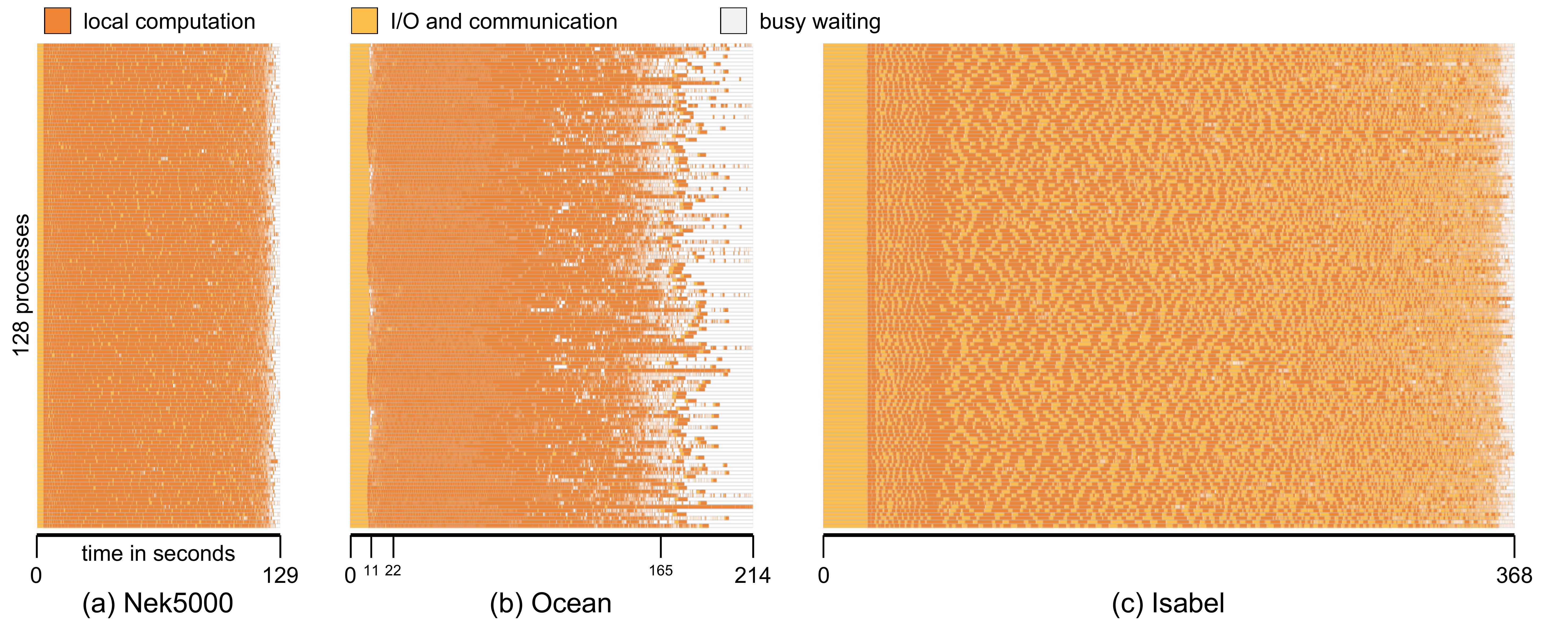}
  \caption{
  Gantt charts for our method using $128$ processes on (a) Nek5000, (b) Ocean, and (c) Isabel data. Each row of the vertical axis corresponds to a process. The horizontal axis encodes the execution time. }
  \label{fig:gantt_chart}
\end{figure*}

\begin{figure*}[bt]
  \centering 
  \includegraphics[width=\linewidth]{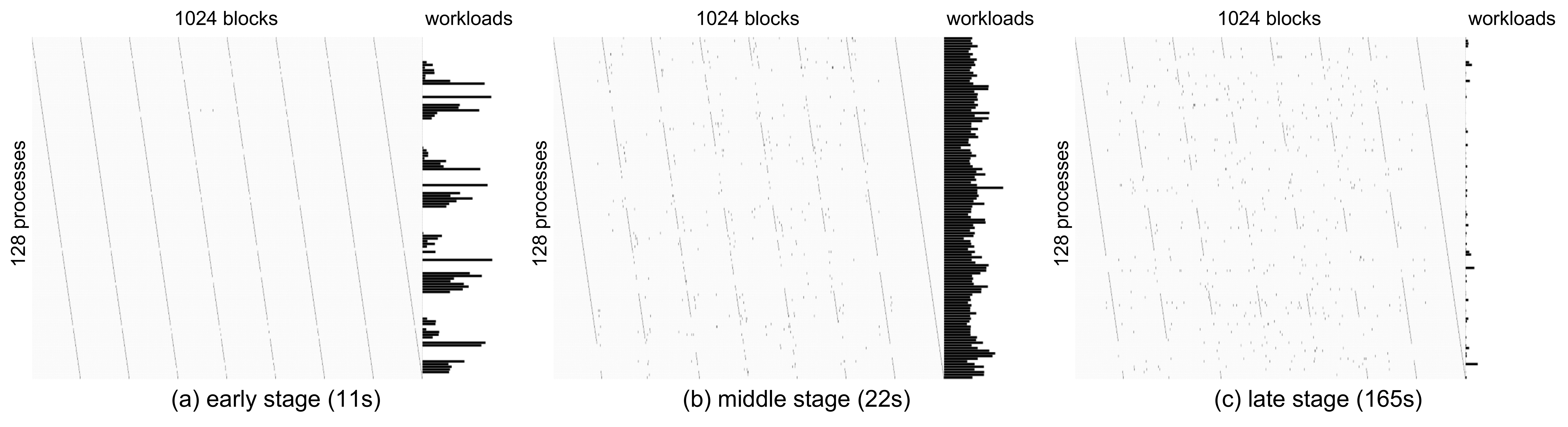}
  \caption{
  For Ocean data, we present time-varying block assignment when using our method with $128$ processes. We extract three snapshots at 11, 22, and 165 seconds to illustrate the change of block assignment change at (a) early, (b) middle, and (c) late stages, respectively, which are also labeled in Fig.~\ref{fig:gantt_chart}b. 
  Each snapshot has two views. First, the left view illustrates the block-to-process assignment, where each row of the vertical axis corresponds to a process, and each column of the horizontal axis corresponds to a block. A dark dot represents a block is assigned to a process at the specified execution time. Second, the right view, a bar chart, encodes the workload of each process using bar length. }
  \label{fig:dynamic_block_assignment}
\end{figure*}

\textbf{Study on per-process performance:} 
To breakdown the performance of our method in detail, we 
profiled activities of each process, including local computation, I/O and communication, and busy waiting in Fig.~\ref{fig:gantt_chart} and supplemental videos. The results in Fig.~\ref{fig:gantt_chart} show that each process has a similar aggregated active time, indicating our method balanced workloads among processes effectively. 
Specific for Ocean data, Fig.~\ref{fig:dynamic_block_assignment} illustrates processes' dynamic block assignment. As shown in Fig.~\ref{fig:dynamic_block_assignment}, the initial workloads of processes were not balanced at the early stage because of local seeding patterns, while the initial block assignment presents the round-robin (a.k.a, block-cyclic) assignment pattern. Our method gradually balanced the processes' workloads by transferring blocks among processes in the middle stage. At the late stage, only a few blocks had particles, making just a few processes had advection computations. 

\begin{figure}[htb]
  \centering 
  \includegraphics[width=\columnwidth]{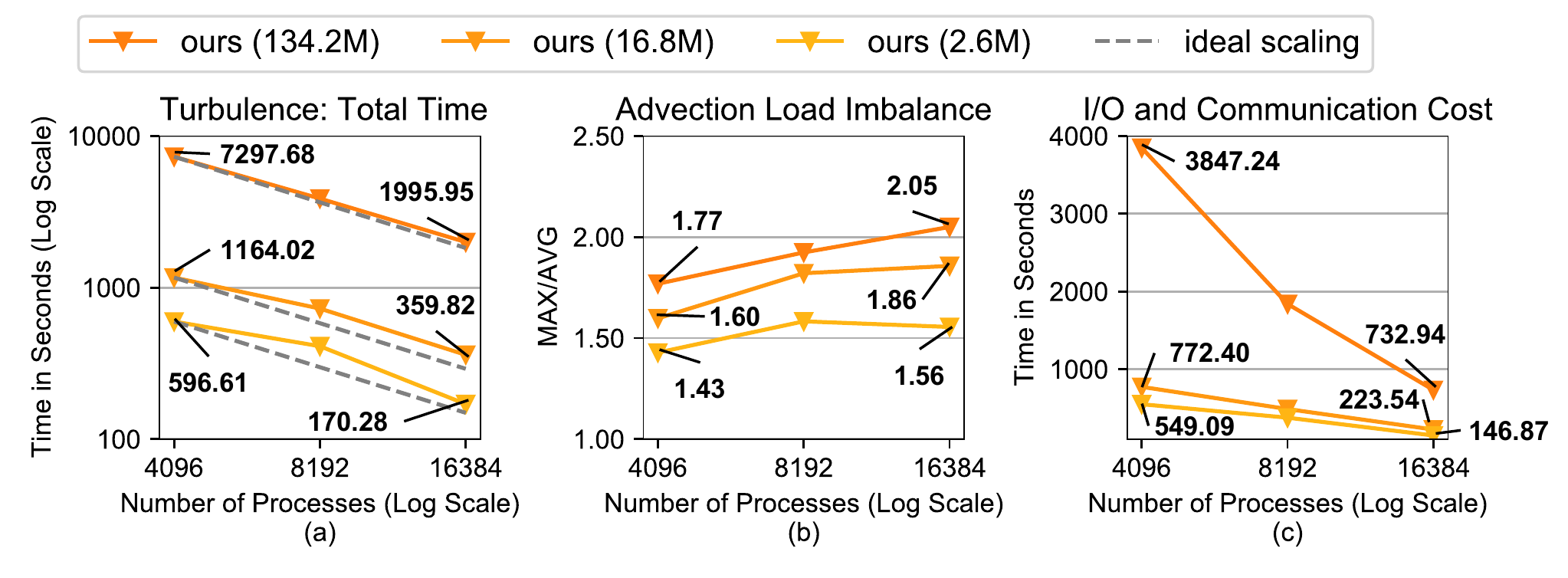}
  \caption{For $4,096^3$ Turbulence data, we present the total execution time, load balance, and I/O and communication time in the three columns separately with $2.6$, $16.8$, and $134.2$ million particles. We evaluate our method from $4,096$ processes to $16,384$ processes on Theta supercomputer. 
  The baseline approach is not presented here due to its execution time exceeding Theta's three-hour execution time constraint.
  }
  \label{fig:strong_scaling_isotropic}
\end{figure}

\textbf{Study on $4,096^3$ Turbulence data:}
We evaluated our method with up to $134.2$ million particle seeds and $16,384$ processes. Turbulence data has the highest spatial resolution and the largest data size among all tested data of the existing parallel particle tracing studies. 

Fig.~\ref{fig:strong_scaling_isotropic} presents the evaluations with varying numbers of particles on Turbulence data. 
Our method remained high strong-scaling efficiencies for varying numbers of particles, which are $87.59\%$ for $2.6$ million particles, $80.88\%$ for $16.8$ million particles, and $91.41\%$ for $134.2$ million particles. The advection load imbalance curve increased as the number of particles grew. 
The I/O and communication cost decreased quickly with respect to the increment of processes. Our method's overheads were $10.65$ seconds for $2.6$ million particles, $14.90$ seconds for $16.8$ million particles, and $27.20$ seconds for $134.2$ million particles when $16,384$ processes were used, which were $6.26\%$, $4.14\%$, and $1.36\%$ of the total execution time, respectively.

\section{Discussions} 
\label{sect:rl_discussion}

\textbf{Optimality limitations:}
We used the estimates of workloads to compute rewards for agents. The quality of discovered strategy depends on the estimation. If estimation errors are high, even if the agents can learn an optimal policy, the workload balance may still be suboptimal. To remedy this, in the future, we are going to train agents to calibrate and adjust the estimates of workloads by computing loss using the difference between the estimates and the real computation time, given the entry point of a particle and the vector field block, which may further decrease the estimation errors. 

\textbf{Out-of-core support: }
If the distributed memory of participating processes is not enough to hold the newly generated data (e.g., for in situ computation), we use the least-recently-used (LRU) rule to evict data blocks having no particles to save space for new data, following Pugmire et al.~\cite{pugmire2009scalable}. 
Because evicted data blocks may be reloaded in the future, we also write constructed trajectories trees of the blocks to disk for possible future reuse. 

\textbf{Trade off between I/O and communications:}
Our algorithm currently is designed to favor migrating data blocks among processes rather than loading data blocks from disks because most HPC clusters and supercomputers have network bandwidth higher than I/O, which may not be beneficial for a computing platform with low network bandwidth and high-speed I/O. In the future, we will incorporate the estimation of I/O costs into our cost model to allow agents to fetch data blocks from disks if I/O costs are lower.

\section{Conclusion and Future Works}
We exploit the benefits of using RL-based optimization to address a dynamic workload balancing and communication cost reduction problem for parallel flow visualization and analysis. The reinforcement learning has a dynamic nature and takes advantage of reward and cost functions to balance workloads and minimize communication costs. An optimized workload balance is achieved by considering communication costs for data transfer, enhancing the parallel particle tracing performance. We evaluate our approach with fluid dynamics, ocean simulation, and weather simulation data and analyze our method's workload balance and scaling efficiency performance. The results demonstrate that our RL-based technique can dynamically make block assignment decisions to minimize the total execution time. 

\textbf{Future works:}
We identify four potential future works. 
First, we can allow agents to both donate and steal works from other processes. The work donation and stealing schemes can complement each other. For example, when most processes have high workloads and just a few are idle, those idle processes' agents can efficiently steal works from others rather than waiting for donations. 
Second, we can allow duplicating data blocks, which can further improve the performance when just a few blocks have particles, for example, at the late stage in Fig.~\ref{fig:dynamic_block_assignment}. 
Third, we can use deep neural networks instead of single-layer perceptrons to formulate latent functions, which may exploit additional learnability for agents. 
Fourth, we will improve and evaluate our algorithm on unsteady flow for in situ computation.




%



\ifCLASSOPTIONcompsoc
  \section*{Acknowledgments}
\else
  \section*{Acknowledgment}
\fi
This work is supported in part by the National Science Foundation Division of Information and Intelligent Systems-1955764, the National Science Foundation Office of Advanced Cyberinfrastructure-2112606, U.S. Department of Energy Los Alamos National Laboratory contract 47145, and UT-Battelle LLC contract 4000159447 program manager Margaret Lentz. 
This research is also supported by the Exascale Computing Project (ECP), project number 17-SC-20-SC, a collaborative effort of the U.S. Department of Energy Office of Science and the National Nuclear Security Administration, as part of the Co-design center for Online Data Analysis and Reduction (CODAR). It is also supported by the U.S. Department of Energy, Office of Advanced Scientific Computing Research, Scientific Discovery through Advanced Computing (SciDAC) program, and by Laboratory Directed Research and Development (LDRD) funding from Argonne National Laboratory, provided by the Director, Office of Science, of the U.S. Department of Energy under Contract No. DE-AC02-06CH11357. 




\ifCLASSOPTIONcaptionsoff
  \newpage
\fi



\bibliographystyle{IEEEtran}
\bibliography{load_balancing}

\begin{thebibliography}{10}
\providecommand{\url}[1]{#1}
\csname url@samestyle\endcsname
\providecommand{\newblock}{\relax}
\providecommand{\bibinfo}[2]{#2}
\providecommand{\BIBentrySTDinterwordspacing}{\spaceskip=0pt\relax}
\providecommand{\BIBentryALTinterwordstretchfactor}{4}
\providecommand{\BIBentryALTinterwordspacing}{\spaceskip=\fontdimen2\font plus
\BIBentryALTinterwordstretchfactor\fontdimen3\font minus
  \fontdimen4\font\relax}
\providecommand{\BIBforeignlanguage}[2]{{%
\expandafter\ifx\csname l@#1\endcsname\relax
\typeout{** WARNING: IEEEtran.bst: No hyphenation pattern has been}%
\typeout{** loaded for the language `#1'. Using the pattern for}%
\typeout{** the default language instead.}%
\else
\language=\csname l@#1\endcsname
\fi
#2}}
\providecommand{\BIBdecl}{\relax}
\BIBdecl

\bibitem{cabral1993imaging}
B.~Cabral and L.~C. Leedom, ``Imaging vector fields using line integral
  convolution,'' in \emph{Proc. Annual Conference on Computer Graphics and
  Interactive Techniques}, 1993, pp. 263--270.

\bibitem{muraki2003pc}
S.~Muraki, E.~B. Lum, K.-L. Ma, M.~Ogata, and X.~Liu, ``A {PC} cluster system
  for simultaneous interactive volumetric modeling and visualization,'' in
  \emph{Proc. IEEE Symposium on Parallel and Large-Data Visualization and
  Graphics}, 2003, pp. 95--102.

\bibitem{haller2001distinguished}
G.~Haller, ``Distinguished material surfaces and coherent structures in
  three-dimensional fluid flows,'' \emph{Physica D: Nonlinear Phenomena}, vol.
  149, no.~4, pp. 248--277, 2001.

\bibitem{nouanesengsy2012parallel}
B.~Nouanesengsy, T.-Y. Lee, K.~Lu, H.-W. Shen, and T.~Peterka, ``{Parallel
  particle advection and FTLE computation for time-varying flow fields},'' in
  \emph{Proc. International Conference on High Performance Computing,
  Networking, Storage and Analysis}, 2012, pp. 61:1--11.

\bibitem{zhang2017dynamic}
J.~Zhang, H.~Guo, F.~Hong, X.~Yuan, and T.~Peterka, ``Dynamic load balancing
  based on constrained kd tree decomposition for parallel particle tracing,''
  \emph{IEEE Transactions on Visualization and Computer Graphics}, vol.~24,
  no.~1, pp. 954--963, 2017.

\bibitem{lu2014scalable}
K.~Lu, H.-W. Shen, and T.~Peterka, ``Scalable computation of stream surfaces on
  large scale vector fields,'' in \emph{Proc. International Conference for High
  Performance Computing, Networking, Storage and Analysis}, 2014, pp.
  1008--1019.

\bibitem{kendall2011simplified}
W.~Kendall, J.~Wang, M.~Allen, T.~Peterka, J.~Huang, and D.~Erickson,
  ``Simplified parallel domain traversal,'' in \emph{Proc. International
  Conference for High Performance Computing, Networking, Storage and Analysis},
  2011, pp. 10:1--11.

\bibitem{guo2014advection}
H.~Guo, J.~Zhang, R.~Liu, L.~Liu, X.~Yuan, J.~Huang, X.~Meng, and J.~Pan,
  ``Advection-based sparse data management for visualizing unsteady flow,''
  \emph{IEEE Transactions on Visualization and Computer Graphics}, vol.~20,
  no.~12, pp. 2555--2564, 2014.

\bibitem{pugmire2009scalable}
D.~Pugmire, H.~Childs, C.~Garth, S.~Ahern, and G.~H. Weber, ``Scalable
  computation of streamlines on very large datasets,'' in \emph{Proc. of the
  Conference on High Performance Computing Networking, Storage and Analysis},
  2009, pp. 16:1--12.

\bibitem{peterka2011study}
T.~Peterka, R.~Ross, B.~Nouanesengsy, T.-Y. Lee, H.-W. Shen, W.~Kendall, and
  J.~Huang, ``A study of parallel particle tracing for steady-state and
  time-varying flow fields,'' in \emph{Proc. International Parallel \&
  Distributed Processing Symposium}, 2011, pp. 580--591.

\bibitem{garey1974some}
M.~R. Garey, D.~S. Johnson, and L.~Stockmeyer, ``Some simplified {NP}-complete
  problems,'' in \emph{Proc. ACM symposium on Theory of computing}, 1974, pp.
  47--63.

\bibitem{gary1979computers}
M.~R. Gary and D.~S. Johnson, ``Computers and intractability: A guide to the
  theory of {NP}-completeness,'' 1979.

\bibitem{sutton2018reinforcement}
R.~S. Sutton and A.~G. Barto, \emph{Reinforcement learning: An
  introduction}.\hskip 1em plus 0.5em minus 0.4em\relax MIT press, 2018.

\bibitem{kielmann2001network}
T.~Kielmann, H.~E. Bal, S.~Gorlatch, K.~Verstoep, and R.~F. Hofman, ``Network
  performance-aware collective communication for clustered wide-area systems,''
  \emph{Parallel Computing}, vol.~27, no.~11, pp. 1431--1456, 2001.

\bibitem{chan2007collective}
E.~Chan, M.~Heimlich, A.~Purkayastha, and R.~Van De~Geijn, ``Collective
  communication: Theory, practice, and experience,'' \emph{Concurrency and
  Computation: Practice and Experience}, vol.~19, no.~13, pp. 1749--1783, 2007.

\bibitem{traff2019optimal}
J.~L. Tr{\"a}ff, ``On optimal trees for irregular gather and scatter
  collectives,'' \emph{IEEE Transactions on Parallel and Distributed Systems},
  vol.~30, no.~9, pp. 2060--2074, 2019.

\bibitem{sujudi1996integration}
D.~Sujudi and R.~Haimes, ``Integration of particle paths and streamlines in a
  spatially-decomposed computation,'' in \emph{Parallel Computational Fluid
  Dynamics 1995}, A.~Ecer, J.~Periaux, N.~Satdfuka, and S.~Taylor, Eds.\hskip
  1em plus 0.5em minus 0.4em\relax Elsevier, 1996, pp. 315--322.

\bibitem{yu2007parallel}
H.~Yu, C.~Wang, and K.-L. Ma, ``{Parallel hierarchical visualization of large
  time-varying 3D vector fields},'' in \emph{Proc. ACM/IEEE Conference on
  Supercomputing}, 2007, pp. 24:1--12.

\bibitem{chen2008optimizing}
L.~Chen and I.~Fujishiro, ``Optimizing parallel performance of streamline
  visualization for large distributed flow datasets,'' in \emph{Proc. Pacific
  Visualization Symposium}, 2008, pp. 87--94.

\bibitem{nouanesengsy2011load}
B.~Nouanesengsy, T.-Y. Lee, and H.-W. Shen, ``Load-balanced parallel streamline
  generation on large scale vector fields,'' \emph{IEEE Transactions on
  Visualization and Computer Graphics}, vol.~17, no.~12, pp. 1785--1794, 2011.

\bibitem{muller2013distributed}
C.~M{\"u}ller, D.~Camp, B.~Hentschel, and C.~Garth, ``Distributed parallel
  particle advection using work requesting,'' in \emph{Proc. IEEE Symposium on
  Large-Scale Data Analysis and Visualization}, 2013, pp. 1--6.

\bibitem{guo2013coupled}
H.~Guo, X.~Yuan, J.~Huang, and X.~Zhu, ``Coupled ensemble flow line advection
  and analysis,'' \emph{IEEE Transactions on Visualization and Computer
  Graphics}, vol.~19, no.~12, pp. 2733--2742, 2013.

\bibitem{guo2014scalable}
H.~Guo, F.~Hong, Q.~Shu, J.~Zhang, J.~Huang, and X.~Yuan, ``Scalable
  {L}agrangian-based attribute space projection for multivariate unsteady flow
  data,'' in \emph{Proc. IEEE Pacific Visualization Symposium}, 2014, pp.
  33--40.

\bibitem{zhang2018dynamic}
J.~Zhang, H.~Guo, X.~Yuan, and T.~Peterka, ``Dynamic data repartitioning for
  load-balanced parallel particle tracing,'' in \emph{Proc. IEEE Pacific
  Visualization Symposium (PacificVis)}, 2018, pp. 86--95.

\bibitem{binyahib2019lifeline}
R.~Binyahib, D.~Pugmire, B.~Norris, and H.~Childs, ``A lifeline-based approach
  for work requesting and parallel particle advection,'' in \emph{Proc.
  Symposium on Large Data Analysis and Visualization}.\hskip 1em plus 0.5em
  minus 0.4em\relax IEEE, 2019, pp. 52--61.

\bibitem{camp2013gpu}
D.~Camp, H.~Krishnan, D.~Pugmire, C.~Garth, I.~Johnson, E.~W. Bethel, K.~I.
  Joy, and H.~Childs, ``{GPU Acceleration of Particle Advection Workloads in a
  Parallel, Distributed Memory Setting},'' in \emph{Eurographics Symposium on
  Parallel Graphics and Visualization}, F.~Marton and K.~Moreland, Eds.\hskip
  1em plus 0.5em minus 0.4em\relax The Eurographics Association, 2013.

\bibitem{childs2014particle}
H.~Childs, S.~Biersdorff, D.~Poliakoff, D.~Camp, and A.~D. Malony, ``Particle
  advection performance over varied architectures and workloads,'' in
  \emph{Proc. International Conference on High Performance Computing}, 2014,
  pp. 1--10.

\bibitem{lane1994ufat}
D.~A. Lane, ``{UFAT}-a particle tracer for time-dependent flow fields,'' in
  \emph{Proc. Visualization}, 1994, pp. 257--264.

\bibitem{lane1995parallelizing}
------, ``Parallelizing a particle tracer for flow visualization,'' Society for
  Industrial and Applied Mathematics, Philadelphia, PA (United States), Tech.
  Rep., 1995.

\bibitem{cabral1995highly}
B.~Cabral and L.~C. Leedom, ``Highly parallel vector visualization using line
  integral convolution,'' in \emph{Proc. SIAM Conference on Parallel Processing
  for Scientific Computing, (PPSC)}, 1995, pp. 802--807.

\bibitem{camp2010streamline}
D.~Camp, C.~Garth, H.~Childs, D.~Pugmire, and K.~Joy, ``Streamline integration
  using {MPI}-hybrid parallelism on a large multicore architecture,''
  \emph{IEEE Transactions on Visualization and Computer Graphics}, vol.~17,
  no.~11, pp. 1702--1713, 2010.

\bibitem{pugmire2018performance}
D.~Pugmire, A.~Yenpure, M.~Kim, J.~Kress, R.~Maynard, H.~Childs, and
  B.~Hentschel, ``{Performance-Portable Particle Advection with VTK-m},'' in
  \emph{Proc. Eurographics Symposium on Parallel Graphics and Visualization},
  2018.

\bibitem{schwartz2021machine}
S.~D. Schwartz, H.~Childs, and D.~Pugmire, ``Machine learning-based autotuning
  for parallel particle advection,'' in \emph{Proc. Eurographics Symposium on
  Parallel Graphics and Visualization}, 2021.

\bibitem{pugmire2012parallel}
D.~Pugmire, T.~Peterka, and C.~Garth, ``Parallel integral curves,'' in
  \emph{High Performance Visualization: Enabling Extreme Scale Scientific
  Insight}, C.~H. E.~Wes~Bethel, Hank~Childs, Ed.\hskip 1em plus 0.5em minus
  0.4em\relax CRC Press, 2012, pp. 13--30.

\bibitem{zhang2018survey}
J.~Zhang and X.~Yuan, ``A survey of parallel particle tracing algorithms in
  flow visualization,'' \emph{Journal of Visualization}, vol.~21, no.~3, pp.
  351--368, 2018.

\bibitem{berger1987partitioning}
M.~J. Berger and S.~H. Bokhari, ``A partitioning strategy for nonuniform
  problems on multiprocessors,'' \emph{IEEE Transactions on Computers}, no.~5,
  pp. 570--580, 1987.

\bibitem{blumofe1999scheduling}
R.~D. Blumofe and C.~E. Leiserson, ``Scheduling multithreaded computations by
  work stealing,'' \emph{Journal of the ACM (JACM)}, vol.~46, no.~5, pp.
  720--748, 1999.

\bibitem{dinan2009scalable}
J.~Dinan, D.~B. Larkins, P.~Sadayappan, S.~Krishnamoorthy, and J.~Nieplocha,
  ``Scalable work stealing,'' in \emph{Proc. Conference on High Performance
  Computing Networking, Storage and Analysis}, 2009, pp. 53:1--11.

\bibitem{saraswat2011lifeline}
V.~A. Saraswat, P.~Kambadur, S.~Kodali, D.~Grove, and S.~Krishnamoorthy,
  ``Lifeline-based global load balancing,'' \emph{ACM SIGPLAN Notices},
  vol.~46, no.~8, pp. 201--212, 2011.

\bibitem{xu2010flow}
L.~Xu and H.-W. Shen, ``{Flow Web: a graph based user interface for 3D flow
  field exploration},'' in \emph{Visualization and Data Analysis}, vol. 7530,
  2010, p. 75300F.

\bibitem{chen2011flow}
C.-M. Chen, L.~Xu, T.-Y. Lee, and H.-W. Shen, ``A flow-guided file layout for
  out-of-core streamline computation,'' in \emph{Proc. IEEE Symposium on Large
  Data Analysis and Visualization}, 2011, pp. 115--116.

\bibitem{chen2012flow}
C.-M. Chen, B.~Nouanesengsy, T.-Y. Lee, and H.-W. Shen, ``Flow-guided file
  layout for out-of-core pathline computation,'' in \emph{Proc. IEEE Symposium
  on Large Data Analysis and Visualization}, 2012, pp. 109--112.

\bibitem{zhang2016efficient}
J.~Zhang, H.~Guo, and X.~Yuan, ``Efficient unsteady flow visualization with
  high-order access dependencies,'' in \emph{Proc. IEEE Pacific Visualization
  Symposium (PacificVis)}, 2016, pp. 80--87.

\bibitem{hong2018access}
F.~Hong, J.~Zhang, and X.~Yuan, ``Access pattern learning with long short-term
  memory for parallel particle tracing,'' in \emph{Proc. IEEE Pacific
  Visualization Symposium}, 2018, pp. 76--85.

\bibitem{tricoche2002topology}
X.~Tricoche, T.~Wischgoll, G.~Scheuermann, and H.~Hagen, ``Topology tracking
  for the visualization of time-dependent two-dimensional flows,''
  \emph{Computers \& Graphics}, vol.~26, no.~2, pp. 249--257, 2002.

\bibitem{garth2004tracking}
C.~Garth, X.~Tricoche, and G.~Scheuermann, ``Tracking of vector field
  singularities in unstructured 3{D} time-dependent datasets,'' in \emph{Proc.
  IEEE Visualization}, 2004, pp. 329--336.

\bibitem{atkeson1997comparison}
C.~G. Atkeson and J.~C. Santamaria, ``A comparison of direct and model-based
  reinforcement learning,'' in \emph{Proc. International Conference on Robotics
  and Automation}, vol.~4, 1997, pp. 3557--3564.

\bibitem{polydoros2017survey}
A.~S. Polydoros and L.~Nalpantidis, ``Survey of model-based reinforcement
  learning: {A}pplications on robotics,'' \emph{Journal of Intelligent \&
  Robotic Systems}, vol.~86, no.~2, pp. 153--173, 2017.

\bibitem{kaiser2019model}
L.~Kaiser, M.~Babaeizadeh, P.~Milos, B.~Osinski, R.~H. Campbell, K.~Czechowski,
  D.~Erhan, C.~Finn, P.~Kozakowski, S.~Levine, A.~Mohiuddin, R.~Sepassi,
  G.~Tucker, and H.~Michalewski, ``Model-based reinforcement learning for
  atari,'' \emph{arXiv preprint arXiv:1903.00374}, 2019.

\bibitem{pal2020brief}
C.-V. Pal and F.~Leon, ``Brief survey of model-based reinforcement learning
  techniques,'' in \emph{Proc. International Conference on System Theory,
  Control and Computing}, 2020, pp. 92--97.

\bibitem{moerland2020model}
T.~M. Moerland, J.~Broekens, and C.~M. Jonker, ``Model-based reinforcement
  learning: A survey,'' \emph{arXiv preprint arXiv:2006.16712}, 2020.

\bibitem{williams1992simple}
R.~J. Williams, ``Simple statistical gradient-following algorithms for
  connectionist reinforcement learning,'' \emph{Machine learning}, vol.~8, no.
  3-4, pp. 229--256, 1992.

\bibitem{agarwal2019theory}
A.~Agarwal, S.~M. Kakade, J.~D. Lee, and G.~Mahajan, ``On the theory of policy
  gradient methods: Optimality, approximation, and distribution shift,''
  \emph{arXiv preprint arXiv:1908.00261}, 2019.

\bibitem{agarwal2020optimality}
------, ``Optimality and approximation with policy gradient methods in markov
  decision processes,'' in \emph{Proc. Conference on Learning Theory}, 2020,
  pp. 64--66.

\bibitem{mei2020global}
J.~Mei, C.~Xiao, C.~Szepesvari, and D.~Schuurmans, ``On the global convergence
  rates of softmax policy gradient methods,'' in \emph{Proc. International
  Conference on Machine Learning}, 2020, pp. 6820--6829.

\bibitem{glascher2010states}
J.~Gl{\"a}scher, N.~Daw, P.~Dayan, and J.~P. O'Doherty, ``States versus
  rewards: dissociable neural prediction error signals underlying model-based
  and model-free reinforcement learning,'' \emph{Neuron}, vol.~66, no.~4, pp.
  585--595, 2010.

\bibitem{reverdy2015parameter}
P.~Reverdy and N.~E. Leonard, ``Parameter estimation in softmax decision-making
  models with linear objective functions,'' \emph{IEEE Transactions on
  Automation Science and Engineering}, vol.~13, no.~1, pp. 54--67, 2015.

\bibitem{raftery1985model}
A.~E. Raftery, ``A model for high-order markov chains,'' \emph{Journal of the
  Royal Statistical Society: Series B (Methodological)}, vol.~47, no.~3, pp.
  528--539, 1985.

\bibitem{fischer2008petascale}
P.~Fischer, J.~Lottes, D.~Pointer, and A.~Siegel, ``Petascale algorithms for
  reactor hydrodynamics,'' in \emph{Proc. Journal of Physics: Conference
  Series}, vol. 125, no.~1, 2008, p. 012076.

\bibitem{maltrud2005eddy}
M.~E. Maltrud and J.~L. McClean, ``An eddy resolving global 1/10 ocean
  simulation,'' \emph{Ocean Modelling}, vol.~8, no. 1-2, pp. 31--54, 2005.

\bibitem{yeung2012dissipation}
P.~Yeung, D.~Donzis, and K.~Sreenivasan, ``Dissipation, enstrophy and pressure
  statistics in turbulence simulations at high reynolds numbers,''
  \emph{Journal of Fluid Mechanics}, vol. 700, pp. 5--15, 2012.

\bibitem{li2008public}
Y.~Li, E.~Perlman, M.~Wan, Y.~Yang, C.~Meneveau, R.~Burns, S.~Chen, A.~Szalay,
  and G.~Eyink, ``A public turbulence database cluster and applications to
  study lagrangian evolution of velocity increments in turbulence,''
  \emph{Journal of Turbulence}, no.~9, p. N31, 2008.

\bibitem{van2009python}
G.~Van~Rossum and F.~L. Drake, \emph{Python 3 Reference Manual}.\hskip 1em plus
  0.5em minus 0.4em\relax Scotts Valley, CA: CreateSpace, 2009.

\bibitem{ritchie1988c}
D.~M. Ritchie, B.~W. Kernighan, and M.~E. Lesk, \emph{The C programming
  language}.\hskip 1em plus 0.5em minus 0.4em\relax Prentice Hall Englewood
  Cliffs, 1988.

\bibitem{behnel2010cython}
S.~Behnel, R.~Bradshaw, C.~Citro, L.~Dalcin, D.~S. Seljebotn, and K.~Smith,
  ``Cython: The best of both worlds,'' \emph{Computing in Science \&
  Engineering}, vol.~13, no.~2, pp. 31--39, 2010.

\bibitem{dalcin2005mpi}
L.~Dalc{\'\i}n, R.~Paz, and M.~Storti, ``{MPI for Python},'' \emph{Journal of
  Parallel and Distributed Computing}, vol.~65, no.~9, pp. 1108--1115, 2005.

\bibitem{peterka2011scalable}
T.~Peterka, R.~Ross, A.~Gyulassy, V.~Pascucci, W.~Kendall, H.-W. Shen, T.-Y.
  Lee, and A.~Chaudhuri, ``Scalable parallel building blocks for custom data
  analysis,'' in \emph{Proc. IEEE Symposium on Large Data Analysis and
  Visualization}, 2011, pp. 105--112.

\bibitem{morozov2016block}
D.~Morozov and T.~Peterka, ``Block-parallel data analysis with {DIY2},'' in
  \emph{Proc. IEEE Symposium on Large Data Analysis and Visualization}.\hskip
  1em plus 0.5em minus 0.4em\relax IEEE, 2016, pp. 29--36.

\bibitem{morozov2021diy}
------, \emph{{DIY}: data-parallel out-of-core library}, accessed January 2021,
  \url{https://github.com/diatomic/diy}.

\bibitem{kendall2011toward}
W.~Kendall, J.~Huang, T.~Peterka, R.~Latham, and R.~Ross, ``Toward a general
  {I/O} layer for parallel-visualization applications,'' \emph{IEEE Computer
  Graphics and Applications}, vol.~31, no.~6, pp. 6--10, 2011.

\bibitem{paszke2019pytorch}
A.~Paszke, S.~Gross, F.~Massa, A.~Lerer, J.~Bradbury, G.~Chanan, T.~Killeen,
  Z.~Lin, N.~Gimelshein, L.~Antiga \emph{et~al.}, ``Pytorch: An imperative
  style, high-performance deep learning library,'' vol.~32, pp. 8026--8037,
  2019.

\bibitem{paszke2017automatic}
A.~Paszke, S.~Gross, S.~Chintala, G.~Chanan, E.~Yang, Z.~DeVito, Z.~Lin,
  A.~Desmaison, L.~Antiga, and A.~Lerer, ``Automatic differentiation in
  pytorch,'' in \emph{Proc. NIPS Autodiff Workshop}, 2017.

\bibitem{rumelhart1985learning}
D.~E. Rumelhart, G.~E. Hinton, and R.~J. Williams, ``Learning internal
  representations by error propagation,'' in \emph{Parallel Distributed
  Processing: Explorations in the Microstructure of Cognition: Foundations},
  D.~E. {Rumelhart} and J.~L. {McClelland}, Eds.\hskip 1em plus 0.5em minus
  0.4em\relax MIT Press, 1987, pp. 318--362.

\bibitem{Tieleman2012}
T.~Tieleman and G.~Hinton, ``{Lecture 6.5---RmsProp: Divide the gradient by a
  running average of its recent magnitude},'' COURSERA: Neural Networks for
  Machine Learning, 2012,
  \url{https://www.cs.toronto.edu/~tijmen/csc321/slides/lecture_slides_lec6.pdf}.

\bibitem{kingma2014adam}
D.~P. Kingma and J.~Ba, ``Adam: A method for stochastic optimization,''
  \emph{arXiv}, vol. abs/1412.6980, 2014.

\bibitem{mnih2013playing}
V.~Mnih, K.~Kavukcuoglu, D.~Silver, A.~Graves, I.~Antonoglou, D.~Wierstra, and
  M.~A. Riedmiller, ``Playing atari with deep reinforcement learning,''
  \emph{arXiv}, vol. abs/1312.5602, 2013.

\bibitem{mnih2015human}
V.~Mnih, K.~Kavukcuoglu, D.~Silver, A.~A. Rusu, J.~Veness, M.~G. Bellemare,
  A.~Graves, M.~Riedmiller, A.~K. Fidjeland, G.~Ostrovski \emph{et~al.},
  ``Human-level control through deep reinforcement learning,'' \emph{nature},
  vol. 518, no. 7540, pp. 529--533, 2015.

\bibitem{mnih2016asynchronous}
V.~Mnih, A.~P. Badia, M.~Mirza, A.~Graves, T.~P. Lillicrap, T.~Harley,
  D.~Silver, and K.~Kavukcuoglu, ``Asynchronous methods for deep reinforcement
  learning,'' \emph{arXiv}, vol. abs/1602.01783, 2016.

\end{thebibliography}
\end{document}